\documentclass[pdflatex,sn-basic,Numbered,iicol]{sn-jnl}

\usepackage{amsmath}
\usepackage{amssymb}
\usepackage{mathtools}

\usepackage[mathscr]{eucal} 
\usepackage{bm}             
\usepackage{upgreek}        
\usepackage{siunitx}        
\DeclareSIUnit\mmhg{mmHg}

\usepackage{amsthm}
\theoremstyle{definition}

\usepackage{xcolor}
\definecolor{colUniBwOr}{rgb}{0.929,0.431,0.0} 
\definecolor{colUniBwGr}{RGB}{113,112,114} 
\definecolor{matlabBlue}{rgb}{0 0.4470 0.7410} 
\definecolor{matlabOrange}{rgb}{0.8500 0.3250 0.0980} 
\definecolor{matlabYellow}{rgb}{0.9290 0.6940 0.1250} 
\definecolor{matlabGreen}{rgb}{0.4660 0.6740 0.1880} 

\definecolor{displacmentgreen}{HTML}{00891B}   
\definecolor{forcered}{HTML}{00891B}   

\definecolor{OIblue}{HTML}{0072B2}
\definecolor{OIorange}{HTML}{E69F00}
\definecolor{OIgreen}{HTML}{009E73}
\definecolor{OIverm}{HTML}{D55E00}
\definecolor{OIpurple}{HTML}{CC79A7}
\definecolor{OIred}{HTML}{D55E00}

\definecolor{MCBlue}{HTML}{4682B4}
\definecolor{MCGreyparaview}{HTML}{b0b7c3}

\definecolor{Markeryellow}{HTML}{ffff00}

\definecolor{PresentationBlue}{RGB}{170,184,219}
\definecolor{PresentationOrange}{RGB}{241,193,159}
\definecolor{PresentationTeal}{RGB}{166,204,196}
\definecolor{PresentationRose}{RGB}{224,163,170}

\usepackage{graphicx}
\usepackage{caption}
\usepackage{subcaption}
\usepackage{float}
\usepackage{adjustbox}

\graphicspath{{pictures/}{pictures2/}{pictures2/cross_section_mca45}{/pictures/results-jewab-tensile-test/}}

\usepackage{url}
\hypersetup{colorlinks=true,
  linkcolor=black,
  citecolor=black,
  urlcolor=black,
  bookmarksnumbered=true,
  pdfstartpage=1,
  pdfpagemode=UseOutlines}

\usepackage[capitalize,nameinlink]{cleveref}
\crefname{figure}{Fig.}{Figs.}
\Crefname{figure}{Figure}{Figures}
\crefname{equation}{Eq.}{Eqs.}
\Crefname{equation}{Equation}{Equations}

\makeatletter
\def\ps@pprintTitle{%
  \let\@oddhead\@empty
  \let\@evenhead\@empty
  \def\@oddfoot{\reset@font\hfil\thepage\hfil}
  \let\@evenfoot\@oddfoot
}
\makeatother

\usepackage{tikz}
\usetikzlibrary{shapes, arrows, calc, fit, backgrounds, shapes.multipart}
\usetikzlibrary{positioning}

\usepackage{pgfplots}
\usepackage{pgfplotstable}
\pgfplotsset{compat = newest}
\usepgfplotslibrary{groupplots,fillbetween}

\pgfplotsset{
  myplot/.style={
      width=\linewidth,
      height=0.62\linewidth,
      grid=both,
      major grid style={line width=0.2pt, draw=gray!30},
      minor grid style={line width=0.1pt, draw=gray!15},
      tick align=outside,
      tick style={black!60},
      axis line style={black!70},
      label style={font=\small},
      tick label style={font=\small},
      legend style={font=\small, draw=none, fill=none},
      legend cell align=left,
    }
}

\usepackage{svg}
\usepackage{multirow}
\usepackage{hyphenat}
\usepackage[ruled]{algorithm2e}

\usepackage[acronym]{glossaries}
\setacronymstyle{long-short}
\glsdisablehyper

\newacronym{ac:1D}{1D}{one-dimensional}
\newacronym{ac:2D}{2D}{two-dimensional}
\newacronym{ac:3D}{3D}{three-dimensional}

\newacronym{ac:ALE}{ALE}{Arbitrary Langrangean Eulerian}

\newacronym{ac:MC}{MC}{microcatheter}

\newglossaryentry{ac:CA}{ 
  name={cerebral aneurysm},
  description={cerebral aneurysm},
  plural={CAs},
  first={cerebral aneurysm (CA)},
  firstplural={cerebral aneurysms (CAs)},
  text={CA}
}

\newglossaryentry{ac:ICA}{
  name={intracranial aneurysm},
  description={intracranial aneurysm},
  plural={intracranial aneurysms},
  first={intracranial aneurysm (ICA)},
  firstplural={intracranial aneurysms (ICAs)},
  text={ICA}
}

\newglossaryentry{ac:FD}{
  name={flow diverter},
  description={flow diverter},
  plural={FDs},
  first={flow diverter (FD)},
  firstplural={flow diverters (FDs)},
  text={FD}
}

\newglossaryentry{ac:SAH}{
  name={subarachnoid hemorrhage},
  description={subarachnoid hemorrhage},
  plural={SAHs},
  first={subarachnoid hemorrhage (SAH)},
  firstplural={subarachnoid hemorrhages (SAHs)},
  text={SAH}
}

\newacronym{ac:MCR}{MCR}{metal coverage ratio}

\newacronym{ac:CPP}{CPP}{closest point projection}

\newacronym{ac:abc}{ABC}{All-Angle-Beam-Contact}

\newacronym{ac:CSM}{CSM}{computational solid mechanics}

\newglossaryentry{ac:DOF}{ 
  name={DOF},
  description={degree of freedom},
  plural={DOFs},
  first={degree of freedom (DOF)},
  firstplural={degrees of freedom (DOFs)},
  text={DOF}
}

\newacronym{ac:FBI}{FBI}{fluid\slash{}beam interaction}

\newacronym{ac:FCM}{FCM}{finite cell method}
\newacronym{ac:CFD}{CFD}{Computational Fluid Dynamics}

\newacronym{ac:FEM}{FEM}{finite element method}

\newacronym{ac:FSI}{FSI}{fluid\slash{}solid interaction}

\newacronym{ac:FVM}{FVM}{finite volume method}

\newacronym{ac:GPTS}{GPTS}{Gauss\hyp{}point\hyp{}to\hyp{}segment}

\newacronym{ac:GPU}{GPU}{graphical processing unit}

\newacronym{ac:HPC}{HPC}{high\hyp{}performance computing}

\newacronym{ac:IBM}{IBM}{immersed boundary method}

\newacronym{ac:IGA}{IGA}{isogeometric analysis}

\newacronym{ac:LagMult}{LM}{Lagrange multiplier}

\newacronym{ac:MPI}{MPI}{message passing interface}

\newacronym{ac:NURBS}{NURBS}{Non-Uniform Rational B-Splines}

\newacronym{ac:PDE}{PDE}{partial differential equation}

\newacronym{ac:PINN}{PINN}{physics-informed neural network}

\newacronym{ac:PI}{PI}{principal investigator}

\newacronym{ac:SPAI}{SPAI}{sparse approximate inverse}

\newacronym{ac:VEM}{VEM}{virtual element method}

\newacronym{ac:WEB}{WEB}{Woven EndoBridge}

\newacronym{ac:WK}{WK}{windkessel}

\newacronym{ac:WP}{WP}{work package}

\newacronym{ac:XFEM}{XFEM}{extended finite element method}

\newacronym{ac:PED}{PED}{Pipeline Embolization Device}

\newacronym{ac:WSS}{WSS}{wall shear stress}

\newacronym{ac:CTA}{CTA}{computed tomography angiography}
\newacronym{ac:MRA}{MRA}{magnetic resonance angiography}

\newacronym{ac:VS}{VS}{virtual stenting}

\usepackage{algcompatible}

\newcommand{\SoftwarePackage}[1]{\textsc{#1}} 
\newcommand{\fourc}{\SoftwarePackage{4C}}

\newcommand{\ie}{i.e.,}

\newcommand{\eg}{e.g.,}

\newcommand{\abaqus}{\SoftwarePackage{Abaqus}}
\newcommand{\beamme}{\SoftwarePackage{BeamMe}}

\newcommand{\transpose}[1]{{#1}^T}						
\newcommand{\abs}[1]{\left| {#1} \right|}				

\newcommand{\identitymatrix}{\mat{I}}

\newcommand{\macaulaybrack}[1]{\langle {#1} \rangle}

\newcommand{\area}{A}			
\newcommand{\youngsmod}{E}		
\newcommand{\shearmod}{G}
\newcommand{\poloarmoment}{I}

\newcommand{\areaof}[1]{\area^{#1}}

\newcommand{\klr}[1]{\left( #1 \right)}
\newcommand{\kle}[1]{\left [  #1 \right ]}

\newcommand{\vct}[1]{\boldsymbol{#1}}
\newcommand{\mat}[1]{\boldsymbol{#1}}

\newcommand{\norm}[1]{\|{#1}\|}

\newcommand{\virtual}{\delta}							
\newcommand{\initial}{0}

\newcommand{\force}{\vct{F}}

\newcommand{\displacement}{\vct{u}}					    
\newcommand{\solid}{S}
\newcommand{\xref}{\vct{X}}
\newcommand{\xrefX}{X_1}
\newcommand{\xrefY}{X_2}
\newcommand{\xrefZ}{X_3}

\newcommand{\xcurr}{\vct{x}}
\newcommand{\us}{\vct{u}^\solid}

\newcommand{\external}{ext}						
\newcommand{\work}{W}

\newcommand{\boundarysolid}{\Gamma^\solid}

\newcommand{\domainsolidinitial}{\Omega^{\solid}_{\initial}}

\newcommand{\potential}{\Pi}

\newcommand{\beam}{B}
\newcommand{\beamdomain}{\Omega^{\beam}}				
\newcommand{\boundaryneumanbeam}{\Gamma_N^{\beam}}				
\newcommand{\beamlenght}{l_{\beam}}
\newcommand{\beamarclength}{s}
\newcommand{\beamcenterline}{\vct{r}}
\newcommand{\beamcenterlines}{\vct{r}\klr{\beamarclength}}
\newcommand{\beammatdefmeas}{\vct{\Gamma}}					
\newcommand{\beammatcurvevector}{\vct{\Omega}}				

\newcommand{\beamcrossectionradius}{R}
\newcommand{\beambasevector}{\vct{g}}
\newcommand{\beamspinvector}{\boldsymbol{\Lambda}}
\newcommand{\beammatbasevector}{\unitvec_1}					
\newcommand{\beamstrainenergy}{\potential^{\beam}_{\mathrm{int}}}
\newcommand{\beamdofs}{\vct{q}}

\newcommand{\virtualbeamcenterline}{\virtual\beamcenterline}
\newcommand{\virtualbeamspinvector}{\virtual\boldsymbol{\theta}}

\newcommand{\virtualworkbeamint}{\virtual \work^{\beam}_{\mathrm{int}}}
\newcommand{\virtualworkbeamext}{\virtual \work^{\beam}_{\mathrm{ext}}}
\newcommand{\virtualworkbeamtot}{\virtual \work^{\beam}_{\mathrm{tot}}}
\newcommand{\virtualworktot}{\virtual \work_{\mathrm{tot}}}

\newcommand{\externalbeammoment}{\vct{m}^{\external}}
\newcommand{\externalbeamforce}{\vct{f}^{\external}}
\newcommand{\neumanbeamforce}{\bar{\vct{f}}}
\newcommand{\neumanbeammoment}{\bar{\vct{m}}}
\newcommand{\beaminternalforce}{\vct{n}}
\newcommand{\beaminternalmoment}{\vct{m}}
\newcommand{\beaminternalforcematerial}{\vct{N}}
\newcommand{\beaminternalmomentmaterial}{\vct{M}}

\newcommand{\beamtosolid}{BS}
\newcommand{\beamtobeam}{BB}

\newcommand{\virtualbeamtosolidpotential}{\virtual\potential_{c}^{\beamtosolid}}
\newcommand{\virtualworkbtbcontact}{\virtual \work^{\beamtobeam}_{\btbpoint}}
\newcommand{\virtualworkbtscontact}{\virtual \work^{\beamtosolid}_{c}}

\newcommand{\beamconsitF}{\mat{C}_F}						
\newcommand{\beamconsitM}{\mat{C}_M}						

\newcommand{\beamyoungsmod}{\youngsmod^{\beam}}
\newcommand{\beamshearsmod}{\shearmod^{\beam}}

\newcommand{\gap}{g}
\newcommand{\contactangle}{\alpha_c}

\newcommand{\contactanglecosine}{z}

\newcommand{\gapbtb}{g_{\beamtobeam}}
\newcommand{\contactdistance}{\vct{d}_c}
\newcommand{\contactpointsmaster}{s_{1c}}
\newcommand{\contactpointsslave}{s_{2c}}
\newcommand{\beamarclengthone}{\beamarclength_1}
\newcommand{\beamarclengthtwo}{\beamarclength_2}
\newcommand{\distbtb}{d_{\beamtobeam}}
\newcommand{\penaltyfactor}{\epsilon}

\newcommand{\btbcouplingcondition}{\Gamma_{\beamtobeam}}
\newcommand{\normalforbtbgap}{\vct{n}^{\beamtobeam}}

\newcommand{\historynormalforbtbgap}{\vct{n}_{prev}^{\beamtobeam}}
\newcommand{\enhancedgap}{\bar{g}_{\beamtobeam}}

\newcommand{\beamtangentvector}{\vct{t}}
\newcommand{\beamtangentvectorcontactmaster}{\beamtangentvector_1(\contactpointsmaster)}
\newcommand{\beamtangentvectorcontactslave}{\beamtangentvector_2(\contactpointsslave)}

\newcommand{\btbpoint}{p}
\newcommand{\btbpointpenalty}{\penaltyfactor_{\btbpoint}}
\newcommand{\btbpointpotential}{\potential^{\beamtobeam}_{\btbpoint}}
\newcommand{\btbpointcontactforce}{\force_{c\btbpoint}}

\newcommand{\couplingdistance}{\vct{d}_{ij}}
\newcommand{\couplingset}{\mathcal{C}}
\newcommand{\pointcoupling}{\mathrm{cpl}}
\newcommand{\pointcouplingparam}{\penaltyfactor_{\pointcoupling}}
\newcommand{\couplingpotential}{\potential^{\beamtobeam}_{\pointcoupling}}
\newcommand{\virtualworkcoupling}{\virtual \work^{\beamtobeam}_{\pointcoupling}}

\newcommand{\rotglmatrix}{\mat{Q_{GL}} }
\newcommand{\rotloc}{\textit{loc}}
\newcommand{\rotglob}{\textit{glob}}
\newcommand{\rotcenter}{\vct{x}_c}
\newcommand{\rotnodeposition}{\beamcenterline_i}
\newcommand{\unitvec}{\vct{e}}
\newcommand{\unitvecx}{\vct{e}_x}

\newcommand{\rotunitvec}{\vct{e}}
\newcommand{\rotunitlocx}{\rotunitvec^{\rotloc}_x}
\newcommand{\rotunitlocy}{\rotunitvec^{\rotloc}_y}
\newcommand{\rotunitlocz}{\rotunitvec^{\rotloc}_z}

\newcommand{\rotunitglobx}{\rotunitvec^{\rotglob}_x}

\newcommand{\btscouplingcondition}{\Gamma_{\beamtosolid}}
\newcommand{\solidsurfxi}{\xi}					
\newcommand{\solidsurfeta}{\eta}
\newcommand{\solidsurfxicp}{\xi_c}				
\newcommand{\solidsurfetacp}{\eta_c}
\newcommand{\solidsurfnormal}{\vct{n}^{\solid}}
\newcommand{\distbts}{d}
\newcommand{\distbtsunilateral}{d^{\mathrm{ul}}}
\newcommand{\btspenalty}{\penaltyfactor_{c}}
\newcommand{\btspotential}{\potential^{\beamtosolid}_{c}}

\newcommand{\radius}{r}

\newcommand{\diameter}{d}

\newcommand{\fd}{fd}
\newcommand{\length}{l}
\newcommand{\fdtotalhelixlength}{l_{h}}
\newcommand{\fdbraidingangle}{\alpha_{\fd}}
\newcommand{\fdhelixangle}{\beta_{\fd}}
\newcommand{\rfd}{\radius_{\fd}}
\newcommand{\lfd}{\length_{\fd}}
\newcommand{\dfd}{\diameter_{\fd}}

\newcommand{\rfdwire}{\radius_w}

\newcommand{\nfdturns}{n_{\fd,t}}
\newcommand{\fdinterwoovenfamiliy}{n_{if}}
\newcommand{\fdbradingdirection}{d_{b}}
\newcommand{\fdwavelength}{L_{c}}
\newcommand{\fdw}{w_{\fd}} 

\newcommand{\parametert}{t}

\newcommand{\fdxparametric}{\vct{x}_{\fd}\klr{\parametert}}
\newcommand{\fdwireindex}{k}
\newcommand{\fdxparametricsingle}{\vct{x}_{\fd}\klr{\parametert,\fdbradingdirection,\fdwireindex}}
\newcommand{\fdphaseangle}{\theta}
\newcommand{\fdphaseanglefct}{\fdphaseangle(\parametert,\fdbradingdirection,\fdwireindex)}

\newcommand{\fdinterwoovenamplitude}{A_p}
\newcommand{\fdampt}{\radius_{h}(\parametert)}

\newcommand{\nwire}{n_w}

\newcommand{\mcat}{MC}

\newcommand{\dmcat}{\diameter^{\mcat}}
\newcommand{\dmcatinner}{\diameter^{\mcat}_i}
\newcommand{\dmcatouter}{\diameter^{\mcat}_o}
\newcommand{\rmcatinner}{\radius^{\mcat}_i}

\newcommand{\mcatthickness}{t^{\mcat}}
\newcommand{\dmcatinnercompressed}{\diameter^{\mcat}_c}

\newcommand{\lmcat}{\length_{\mcat}}

\newcommand{\loadstepparameter}{\lambda}

\newcommand{\pressure}{p}

\newcommand{\displacementtensiletest}{\bar{\displacement}_t}
\newcommand{\tensiletestaxialforce}{\vct{F}_{ax}}

\newcommand{\strainmax}{\varepsilon_{\text{max}}}

\newcommand{\displacementcompressiontest}{\bar{\displacement}_c}
\newcommand{\radialdisplacementcompressiontest}{\bar{u}_{c,r}}
\newcommand{\logisticsteepness}{a_{\lambda}}
\newcommand{\logisticaxialrate}{a_x}
\newcommand{\loadstepshift}{\loadstepparameter_{\mathrm{shift}}}
\newcommand{\targetradius}{\radius_{I}}

\newcommand{\segmentindex}{k}
\newcommand{\segmenttargetradius}{\hat{\radius}_{\segmentindex}}
\newcommand{\segmentcenter}{X_{1,\segmentindex}}
\newcommand{\segmentwidth}{\Delta X_{\mathrm{c}}}
\newcommand{\segmenttransitionwidth}{\Delta X_{\mathrm{trans}}}

\newcommand{\mcr}{\mathrm{MCR}}
\newcommand{\mcrporositiy}{\phi}

\begin{document}

\title{Mechanical Modeling of Braided Neurovascular Flow Diverters using a Beam-to-Beam and Beam-to-Surface Contact Formulation}

\author*[1]{\fnm{Martin} \sur{Frank}}\email{martin.frank@unibw.de}
\author[1]{\fnm{Ivo} \sur{Steinbrecher}}\email{ivo.steinbrecher@unibw.de}
\author[1,2]{\fnm{Matthias} \sur{Mayr}}\email{matthias.mayr@unibw.de}
\author[1]{\fnm{Alexander} \sur{Popp}}\email{alexander.popp@unibw.de}

\affil*[1]{\orgdiv{Institute for Mathematics and Computer-Based Simulation}, \orgname{University of the Bundeswehr Munich}, \orgaddress{\street{Werner-Heisenberg-Weg 39}, \city{Neubiberg}, \postcode{85577}, \country{Germany}}}
\affil[2]{\orgdiv{Data Science \& Computing Lab}, \orgname{University of the Bundeswehr Munich}, \orgaddress{\street{Werner-Heisenberg-Weg 39}, \city{Neubiberg}, \postcode{85577}, \country{Germany}}}

\abstract{Braided neurovascular flow diverters are widely used for the endovascular
  treatment of intracranial aneurysms. Their mechanical response and deployed
  configuration are governed by the interaction of many slender, interwoven wires, which
  makes the numerical simulation of the complete device a challenging problem. This work
  presents a detailed mechanical model of complete braided flow diverters that integrates
  geometrically exact beam mechanics and contact mechanics into one consistent implicit
  finite element formulation. Every wire is modeled as a geometrically exact
  Simo--Reissner beam, which consistently describes large rotations and curved reference
  configurations. The interactions between the wires are captured by a beam-to-beam
  contact formulation with a history-dependent gap function that prevents unintended wire
  crossings. The interaction with surrounding tubular structures, such as microcatheters
  or crimping devices, is captured by a beam-to-surface contact formulation. A parametric
  description of interwoven braided geometries enables systematic control of design
  parameters such as wire count, braiding angle, device length, and interweaving pattern.
  The model is assessed by means of three test cases from the literature. A tensile test
  investigates the length--diameter relation and the axial force response, and a radial
  compression test addresses the pressure--diameter behavior. A stepwise compression
  example evaluates geometry-sensitive quantities, including local wire distance, helix
  angle, porosity, and metal coverage ratio. The results show that the model captures both
  the characteristic mechanical response and clinically relevant geometric quantities of
  braided flow diverters.}


\maketitle

\clearpage

\section{Introduction}



Intracranial aneurysms, also referred to as \glspl{ac:CA}, are characterized by a
localized and pathological weakening of the arterial wall, potentially driven by
inflammatory processes~\cite{maragkosOverviewDifferentFlow2020}. This weakening leads
to an abnormal dilation of the vessel and represents a serious cerebrovascular disease.
Rupture of an intracranial aneurysm can result in subarachnoid hemorrhage, which is
associated with high mortality or severe long-term
disability~\cite{brismanCerebralAneurysms2006}. Roughly six out of every \num{100000}
people per year will suffer such a subarachnoid hemorrhage in the general
population~\cite{Rinkel1998}. The prevalence, typical locations, size distributions,
and rupture risks of intracranial aneurysms have been extensively studied and are well
documented in the
literature~\cite{vlakPrevalenceUnrupturedIntracranial2011,wiebersUnrupturedIntracranialAneurysms2003,jeongSizeLocationRuptured2009,wermerRiskRuptureUnruptured2007}.

A variety of treatment strategies exist for \glspl{ac:CA}, and the selection of an
appropriate therapy depends on multiple factors, including aneurysm morphology,
anatomical location, patient-specific risks, and experience of the treating
clinicians~\cite{connollyGuidelinesManagementAneurysmal2012,Toth2018}. Compared to
other interventional techniques such as surgical clipping, flow diversion offers a
minimally invasive alternative that can reduce procedural stress for the
patient~\cite{dursoFlowDiversionIntracranial2011}. Upon implantation, the device
diverts blood flow away from the aneurysm sac, promoting intra-aneurysmal thrombosis
while preserving perfusion of the parent
vessel~\cite{augsburgerEffectFlowDiverter2009}. Unlike intrasaccular devices such as
coils or \gls{ac:WEB} devices, \glspl{ac:FD} avoid direct contact with the interior of
the aneurysm sac since they are deployed within the parent
artery~\cite{lylykCurativeEndovascularReconstruction2009}. In the literature,
\glspl{ac:FD} are often referred to as braided neurovascular stents or self-expanding
braided wire stents. From a technical perspective, a \gls{ac:FD} can be manufactured
using a braiding process~\cite{kimMechanicalModelingSelfexpandable2008}, in which
multiple wires are interwoven at a prescribed braiding angle. As a result, a
\gls{ac:FD} can be described as a cylindrical, self-expanding mesh of thin wires that
is deployed into the parent vessel during endovascular intervention.

The structural properties of the \gls{ac:FD} are crucial to avoid vascular wall injury
during deployment and to ensure a stable, well-adapted configuration after
implantation. Moreover, the final deployed geometry directly influences clinically
relevant geometric measures such as porosity and \gls{ac:MCR}, which serve as key
inputs for subsequent hemodynamic analyses~\cite{mutEffectsFlowDivertingDevice2012}.

Despite their growing clinical relevance and widespread use, the mechanical and
structural behavior of flow diverters still poses many open questions from a modeling
and numerical perspective. Especially for clinical planning and prediction of
\gls{ac:FD} deployment for patient-specific cases, accurate modeling of \glspl{ac:FD}
still remains an open topic of research. To improve the prediction of \gls{ac:FD}
performance and their self-expanding behavior during deployment, accurate mechanical
models that are validated with respect to both structural and geometric properties are
required~\cite{zaccariaModelingBraidedStents2020,kellyComparisonComputationalModelling2019}.


%


A wide range of stents and \gls{ac:FD} designs has been proposed within different
biomedical application fields over the last decade. The present work focuses
exclusively on purely wire-based \glspl{ac:FD} similar to state-of-the-art commercial
devices~\cite{White2023,maragkosOverviewDifferentFlow2020} used in clinical
interventions of \glspl{ac:CA}. Other existing approaches such as duplex
designs~\cite{alherzNumericalFrameworkMechanical2016} accounting for wires with
different strut sizes or devices incorporating additional polymer
coatings~\cite{giuliodoriNumericalModelingBare2021} are beyond the scope of this
contribution.

To realistically prescribe the initial geometry of a \gls{ac:FD}, several parametric
descriptions exist to represent the interwoven or non-interwoven
geometry~\cite{mckennaFiniteElementInvestigation2021,kimMechanicalModelingSelfexpandable2008,ubachsComputationalModelingBraided2023,frankNumericalSimulationEndovascular2024}.
According to geometric open-~or closed-end~point~geometry, the parametric design can be
adapted to match this different design
precisely~\cite{zaccariaModelingBraidedStents2020}.

To account for the structural behavior realistically, two main approaches are
considered within the literature so far. The first approach is based on analytical
descriptions to establish a relation between the changing geometry and the
forces~\cite{jedwabStudyGeometricalMechanical1993,moonAnalyticalModelsPredicting2009,zaccariaAnalyticalMethodsBraided2021}.
On the other hand, finite-element-based models are quite commonly used. Due to their
slenderness, the thin wires are modeled with beam theory.
A larger set of the literature~\cite{kimMechanicalModelingSelfexpandable2008,
mckennaFiniteElementInvestigation2021,ubachsComputationalModelingBraided2023,zaccariaModelingBraidedStents2020,maComputerModelingDeployment2012}
uses beam theory in the explicit solver of the commercial software~{\abaqus}, such as
linear
Timoshenko~\cite{mckennaFiniteElementInvestigation2021,zaccariaModelingBraidedStents2020}
or Euler--Bernoulli theory~\cite{maComputerModelingDeployment2012}. A different beam
formulation is used in the open-source package
of~\cite{bisighiniEndoBeamsjlJuliaFinite2022}, which follows the co-rotational approach
of~\cite{aguirreImplicit3DCorotational2020} and includes a separate frictional contact
formulation between a beam and a rigid surface. There, kinematic constraints are
imposed on a~\gls{ac:FD} to deploy it to a simplified geometry. The study mainly
demonstrates the framework’s capabilities in a simplified setting, rather than
providing a detailed investigation of the \gls{ac:FD} modeling approach or deployment.

For finite-element-based models, a suitable description due the arising contact between
the individual thin wires must be chosen. To avoid a general, complex contact
description for the interaction between the wires, kinematic constraints may be used to
account for the coupling. Papers~\cite{zaccariaModelingBraidedStents2020}
and~\cite{kellyComparisonComputationalModelling2019} compared different kinematic
constraints by connector elements such as joints or hinges. However, both studies
concluded that introducing such kinematic constraints leads to significant deviations
when compared to general contact formulations, thereby recommending the use of full
contact descriptions for accurate predictions.


Another key parameter during the initial design of such a braided stent is the braiding
angle. Its influence has been investigated by several
authors~\cite{ubachsComputationalModelingBraided2023,kimMechanicalModelingSelfexpandable2008,mckennaExperimentalEvaluationMechanics2020}.
Depending on the definition of the braiding angle, it can be concluded that a higher
braiding angle results in decreased axial compressive, but increased radial forces.




The thin wires forming \glspl{ac:FD} are commonly manufactured from Nitinol or
cobalt--chromium alloys~\cite{junFlowDiverterPerformance2024}. In particular, Nitinol
is widely used due to its superelastic behavior, which enables large recoverable
strains and provides the self-expanding capability required for endovascular
deployment. To capture this behavior realistically, several numerical studies have
incorporated superelastic constitutive models for Nitinol wires, often based on
phenomenological descriptions of stress-induced martensitic phase transformation
\cite{auricchioShapememoryAlloysMacromodelling1997,kimMechanicalModelingSelfexpandable2008}.
These experimental findings further emphasize the need for structural models that
simultaneously account for material nonlinearity, geometric evolution, and wire-to-wire
interactions. Compression, tension, bending, and kink tests consistently reveal a
strong dependence on braiding angle, wire diameter, and crossover mechanics, with
characteristic hysteresis arising from a combination of superelasticity and inter-wire
friction~\cite{kimMechanicalModelingSelfexpandable2008,mckennaExperimentalEvaluationMechanics2020}.

Although substantial progress has been made in modeling braided \glspl{ac:FD},
beam-based models commonly rely on general contact algorithms in explicit time stepping
schemes, or replace the contact between the wires with kinematic constraints. To the
authors' best knowledge, a detailed mechanical model of the complete device that
integrates geometrically exact beam mechanics with explicit wire-wire and wire-surface
contact into one consistent formulation has not yet been established. This work
addresses this gap by developing a mechanical model of the complete device with high
fidelity in its beam kinematics and contact interactions. Every wire is resolved as a
geometrically exact Simo--Reissner beam, and the interactions between the wires and
with surrounding structures are captured by explicit beam-to-beam and beam-to-surface
contact formulations. A history-dependent gap function is introduced to detect and
prevent unintended wire crossings. In addition, a flexible parametric description of
the \gls{ac:FD} geometry is presented, allowing the representation of different
braiding patterns and systematic control over key design parameters such as braiding
angle, wire count, and device length. Furthermore, a dedicated end-point treatment
based on kinematic constraints and local coordinate transformations is employed to
obtain a stable open-end configuration and to prevent fraying at the device ends. The
proposed modeling approach is compared to established literature-based experimental
cases~\cite{jedwabStudyGeometricalMechanical1993,shapiroVariablePorosityPipeline2014}
to demonstrate its validity from both structural and geometric perspectives relevant to
device selection and deployment.

The remainder of this manuscript is organized as follows. \cref{sec:modeling} presents
the mathematical foundations, ranging from the continuum formulation of the
geometrically exact Simo--Reissner beam theory in~\cref{sec:beamtheory} to the
beam-to-beam contact formulation in~\cref{sec:beamtobeamcontact} and the
beam-to-surface contact formulation in~\cref{sec:beam-to-surface-contact}.
Then,~\cref{sec:FD-model} presents the structural model of the~\gls{ac:FD} in detail,
beginning with the choice of the formulations in~\cref{sec:FD-model-explanation},
including the parametric description in~\cref{sec:fdparametric}. Furthermore, the
necessary surface interactions are discussed in~\cref{sec:Interaction-explanation}.
Finally, the proposed modeling framework is evaluated against established benchmark
cases from the literature in~\cref{sec:numerical-examples}. The results are compared to
classical tensile and compression experiments
in~\cref{sec:tensile-example,sec:crimping-test} reported
in~\cite{jedwabStudyGeometricalMechanical1993}, demonstrating the ability of the model
to reproduce characteristic mechanical responses of braided stents. A representative
deployment example is presented in~\cref{sec:shapiro-test} and compared with results
from published studies~\cite{shapiroVariablePorosityPipeline2014}, illustrating the
applicability of the proposed approach to further realistic placement scenarios.
Finally,~\cref{sec:conclusion} will summarize the findings and further outline research
objectives.

\section{Mathematical foundations and mechanical modeling framework}\label{sec:modeling}
This section summarizes the formulations on which the~\gls{ac:FD} model
in~\cref{sec:FD-model} is based, including the interaction with a kinematically
prescribed solid.

\subsection{Simo--Reissner beam theory} \label{sec:beamtheory}

The Simo--Reissner beam theory provides a geometrically exact description of
shear-deformable beams undergoing finite displacements and
rotations~\cite{reissnerOnedimensionalFinitestrainBeam1972,simoDynamicsSpaceRods1988,
meierGeometricallyExactBeam2018}. It models the beam as a one-dimensional Cosserat
continuum with independent translational and rotational fields along the beam
centerline. The configuration is described by the centerline
position~$\beamcenterlines\in\mathbb{R}^3$ and an orthonormal director
triad~$\beamspinvector\klr{\beamarclength}=[\beambasevector_1\klr{\beamarclength},\beambasevector_2\klr{\beamarclength},\beambasevector_3\klr{\beamarclength}]~\in~\mathrm{SO}(3)$
attached to the cross-section, where~$\beamarclength\in\beamdomain=[0,\beamlenght]$
denotes the arc-length parameter in the reference configuration,~$\beamdomain$ is the
one-dimensional beam domain, and~$\beamlenght$ is the reference beam length. The
objective deformation measures are introduced in material form as
\begin{align}
  \beammatdefmeas     & = \transpose{\beamspinvector} \beamcenterline' - \beammatbasevector, \\
  \beammatcurvevector & = \transpose{\beamspinvector} \beamspinvector',
\end{align}
where~$(\cdot)'$ denotes differentiation with respect to~$\beamarclength$ and
$\beammatbasevector$ is the material basis vector aligned with the reference
centerline. The vector~$\beammatdefmeas$ represents axial and shear strains,
while~$\beammatcurvevector$ contains torsional and bending strains. Both measures
are invariant under superposed rigid body motions~\cite{crisfieldObjectivityStrainMeasures1999}.
The internal stress state is characterized by the material force and moment resultants
$\beaminternalforcematerial\klr{\beamarclength}$ and~$\beaminternalmomentmaterial\klr{\beamarclength}$, which are energetically conjugate to
$\beammatdefmeas$ and~$\beammatcurvevector$, respectively. For a quasi-static problem,
the strong form of equilibrium in spatial representation reads
\begin{align}
  \beaminternalforce' + \externalbeamforce                                                & = \vct{0}, \\
  \beaminternalmoment' + \beamcenterline' \times \beaminternalforce + \externalbeammoment & = \vct{0}.
\end{align}
Here~$\externalbeamforce$ and~$\externalbeammoment$ denote distributed external
forces and moments per unit reference length, respectively, and
$\beaminternalforce$ and~$\beaminternalmoment$ refer to the internal force and moment
within the current configuration.
The weak form of equilibrium follows from the principle of virtual work. The
admissible variations are given by the virtual displacement~$\virtualbeamcenterline$
and the virtual spin vector~$\virtualbeamspinvector$. The principle of virtual work reads
\begin{equation}
  \virtualworkbeamint - \virtualworkbeamext  = 0
  \quad \forall\,(\virtualbeamcenterline,\virtualbeamspinvector)
\end{equation} where the internal virtual work is given by
\begin{equation}
  \virtualworkbeamint
  =
  \int_{\beamdomain}
  \left(
  \beaminternalforcematerial\cdot\virtual\beammatdefmeas
  +
  \beaminternalmomentmaterial\cdot\virtual\beammatcurvevector
  \right)\,\mathrm{d}\beamarclength \text{.}
\end{equation}
The internal virtual work expression is obtained from a constitutive law that relates
the objective strain measures to the corresponding internal force and moment
resultants. The associated objective variations are
\begin{align}
  \virtual\beammatdefmeas
   & = \transpose{\beamspinvector} \klr{\virtualbeamcenterline'
  + \beamcenterline' \times \virtualbeamspinvector },                \\
  \virtual\beammatcurvevector
   & = \transpose{\beamspinvector }\virtualbeamspinvector'  \text{.}
\end{align}
Furthermore, a linear elastic material response is assumed,
such that the internal energy can be derived
from a strain energy function~$\beamstrainenergy$, formulated with
respect to the objective deformation measures as
\begin{equation}
  \beamstrainenergy
  =
  \tfrac{1}{2}\,
  \int_{\beamdomain}
  \left(
  \transpose{\beammatdefmeas} \beamconsitF \beammatdefmeas
  +
  \transpose{\beammatcurvevector} \beamconsitM \beammatcurvevector
  \right)\,\mathrm{d}\beamarclength \text{.}
\end{equation}
Taking the first variation of the strain energy with respect to the deformation
measures yields the internal force and moment resultants as
\begin{align}
  \beaminternalforcematerial   = \beamconsitF \beammatdefmeas \qquad \text{and} \qquad
  \beaminternalmomentmaterial  = \beamconsitM\beammatcurvevector\text{.}
\end{align}
The respective material constitutive tensors~$\beamconsitF$ and~$\beamconsitM$ result in
a diagonal structure within the material configuration as
\begin{align}
  \beamconsitF
   & =
  \mathrm{diag}\!\left(
  \beamyoungsmod \areaof{\beam},\;
  \beamshearsmod \areaof{\beam}_2,\;
  \beamshearsmod \areaof{\beam}_3
  \right),\text{~and~} \\
  \beamconsitM
   & =
  \mathrm{diag}\!\left(
  \beamshearsmod \poloarmoment_T,\;
  \beamyoungsmod \poloarmoment_2,\;
  \beamyoungsmod \poloarmoment_3
  \right) \text{.}
\end{align}
Therein,~$\beamyoungsmod$ denotes the Young’s modulus,~$\beamshearsmod$ the shear modulus,~$\areaof{\beam}$ the cross-sectional area,
$\areaof{\beam}_2$ and~$\areaof{\beam}_3$ the effective shear areas,~$\poloarmoment_2$ and~$\poloarmoment_3$ the principal
second moments of area, and~$\poloarmoment_T$ the torsional moment of inertia of the beam.
Finally, the external virtual work for a beam reads
\begin{equation}
  \begin{split}
    \virtualworkbeamext
    ={} &
    \int_{\beamdomain}
    \left(
    \externalbeamforce\cdot\virtualbeamcenterline
    +
    \externalbeammoment\cdot\virtualbeamspinvector
    \right)\,\mathrm{d}\beamarclength
    \\
        & +
    \left[
      \neumanbeamforce\cdot\virtualbeamcenterline
      +
      \neumanbeammoment\cdot\virtualbeamspinvector
      \right]_{\boundaryneumanbeam} \text{,}
  \end{split}
\end{equation}
where the Neumann boundary~$\boundaryneumanbeam\subseteq\{0,\beamlenght\}$
denotes the set of beam end points at which the point forces~$\neumanbeamforce$ and point
moments~$\neumanbeammoment$ are prescribed.
Throughout this work, the problem is considered quasi-static. Therefore, all external
loads and prescribed displacements are parametrized by a dimensionless load
parameter~$\loadstepparameter$, which acts as a pseudo-time.


\subsection{Penalty-based beam-to-beam contact formulation}\label{sec:beamtobeamcontact}

\begin{figure}
  \centering
  \begin{minipage}[b]{\linewidth}
    \centering
    \includegraphics[width=\linewidth]{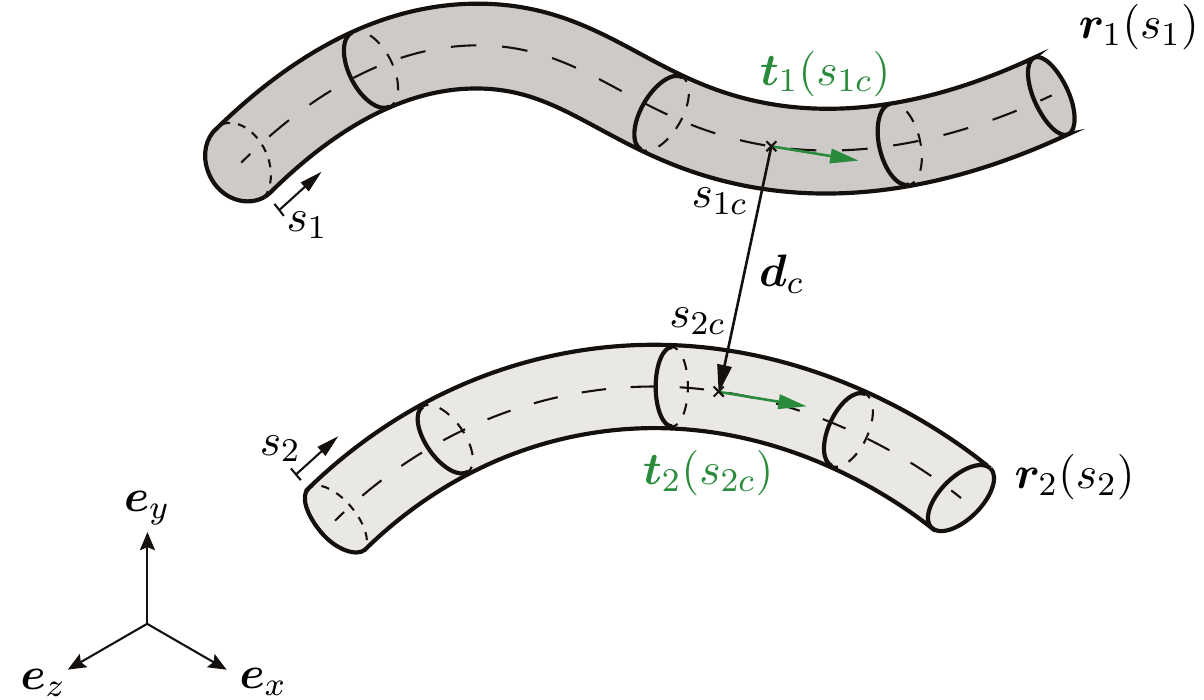}
  \end{minipage}
  \caption{Schematic drawing of a contact scenario for two beams prescribed by the two
    centerlines~$\beamcenterline_1(\beamarclengthone)$ and~$\beamcenterline_2(\beamarclengthtwo)$, with their corresponding tangent vectors~$\beamtangentvector$
    and the distance~$\contactdistance$ obtained from the \acrshort{ac:CPP}.}
  \label{fig:beamcontact}
\end{figure}

To account for the interaction between the individual wires, a suitable mechanical
description of the forces resulting from the contact between individual beams is
required.
Penalty-based contact formulations have been shown to address the problem accurately
and to provide numerically robust
solutions~\cite{meierUnifiedApproachBeamtobeam2017,litewkaFiniteElementAnalysis2010,wriggersContactThreeDimensionalBeams1997}.
Here, the contact between two beams is enforced by a penalty regularization of the
non-penetration condition as displayed in~\cref{fig:beamcontact}. To estimate if and
where precisely the contact between two beam
centerlines~$\beamcenterline_1(\beamarclengthone)$
and~$\beamcenterline_2(\beamarclengthtwo)$ is established or not, the minimization
problem occurring from the \gls{ac:CPP}
\begin{equation}
  (\contactpointsmaster, \contactpointsslave)
  =
  \arg\min_{\beamarclengthone,\beamarclengthtwo}
  \;
  \norm{
    \contactdistance(\beamarclengthone, \beamarclengthtwo)
  }
\end{equation}
must be solved, where the relative distance vector~$\contactdistance$
can be calculated by
\begin{equation}
  \contactdistance(\beamarclengthone, \beamarclengthtwo) = \beamcenterline_1(\beamarclengthone) - \beamcenterline_2(\beamarclengthtwo) \text{.}
\end{equation}
In practice, the \gls{ac:CPP} is computed by solving the orthogonality conditions:
\begin{align}
  \beamtangentvectorcontactmaster \cdot
  \bigl(
  \beamcenterline_1(\contactpointsmaster) - \beamcenterline_2(\contactpointsslave)
  \bigr) & = 0,  \label{eq:cpporthogonality1} \\
  \beamtangentvectorcontactslave \cdot
  \bigl(
  \beamcenterline_1(\contactpointsmaster) - \beamcenterline_2(\contactpointsslave)
  \bigr) & = 0 . \label{eq:cpporthogonality2}
\end{align}
Hence, at the closest point pair the relative distance vector
is orthogonal to both tangent vectors~$\beamtangentvectorcontactmaster$
and~$\beamtangentvectorcontactslave$. This implies that the contact normal direction
can be defined as

\begin{equation}
  \normalforbtbgap
  =
  \frac{\contactdistance}{\norm{\contactdistance}}
  \text{.}
\end{equation}
In general, the nonlinear system~\cref{eq:cpporthogonality1,eq:cpporthogonality2} is
solved numerically, e.g.,\ by Newton's method.
The existence of a unique solution is
assumed for sufficiently large contact angles, while for small contact angles or nearly
parallel beam configurations the \gls{ac:CPP} may become non-unique or ill-conditioned.
The contact angle~$\contactangle \in
  [\qty{0}{\degree},\qty{90}{\degree}]$ between the two beams at the closest point pair is defined by
\begin{equation}
  \contactangle = \arccos\klr{\contactanglecosine}
  \quad \text{with} \quad
  \contactanglecosine =
  \frac{\abs{\beamtangentvectorcontactmaster \cdot \beamtangentvectorcontactslave}}
  {\norm{\beamtangentvectorcontactmaster}\,\norm{\beamtangentvectorcontactslave}}
  \text{.}
\end{equation}
Based on the obtained normal, the norm of the distance~$\distbtb =
\norm{\contactdistance}$ between two beam centerlines and the cross-section
radii~$\beamcrossectionradius_1$,~$\beamcrossectionradius_2$, the gap function is
defined as
\begin{equation}
  \gapbtb = \distbtb - \beamcrossectionradius_1 - \beamcrossectionradius_2 \text{.}
\end{equation}
The \gls{ac:abc} formulation proposed in~\cite{meierUnifiedApproachBeamtobeam2017}
combines a point-contact contribution (as outlined with the \gls{ac:CPP} above) and a
line-contact contribution. In the present work, this formulation is restricted to
the point-contact contribution, yielding a penalty potential of the type introduced
in~\cite{wriggersContactThreeDimensionalBeams1997}. This simplification is justified
for the following examples, since the contact angles between interacting wires
remain comparatively large and the line-contact contribution of the full
\gls{ac:abc} formulation is therefore not activated and the corresponding penalty
potential reads
\begin{equation}
  \begin{split}
    \btbpointpotential
                            & =
    \frac{1}{2}\,\btbpointpenalty\,\macaulaybrack{\gapbtb}^2 \quad \text{~with~}
    \\
    \macaulaybrack{\gapbtb} & = \begin{cases}
                                  \gapbtb & \gapbtb\leq 0 \\
                                  0       & \gapbtb > 0
                                \end{cases} \text{.}
  \end{split}
  \label{eq:btbcontactpoint}
\end{equation}
Here~$\btbpointpenalty$ represents the point contact penalty parameter and
$\macaulaybrack {\cdot}$ ensures that the contact interaction is activated only for~$\gapbtb\leq0$.
The resulting contact force follows as
\begin{equation}
  \btbpointcontactforce
  =
  -\frac{\partial \btbpointpotential}{\partial \beamdofs}
  =
  -\btbpointpenalty\, \macaulaybrack{\gapbtb}
  \frac{\partial \gapbtb}{\partial \beamdofs},
\end{equation}
with~$\beamdofs$ denoting the beam degrees of freedom.
This contribution is computationally efficient and well suited for large contact
angles, but relies on the existence of a unique closest-point projection.

\begin{figure*}
  \centering
  \begin{minipage}[b]{\textwidth}
    \centering
    \includegraphics[width=\linewidth]{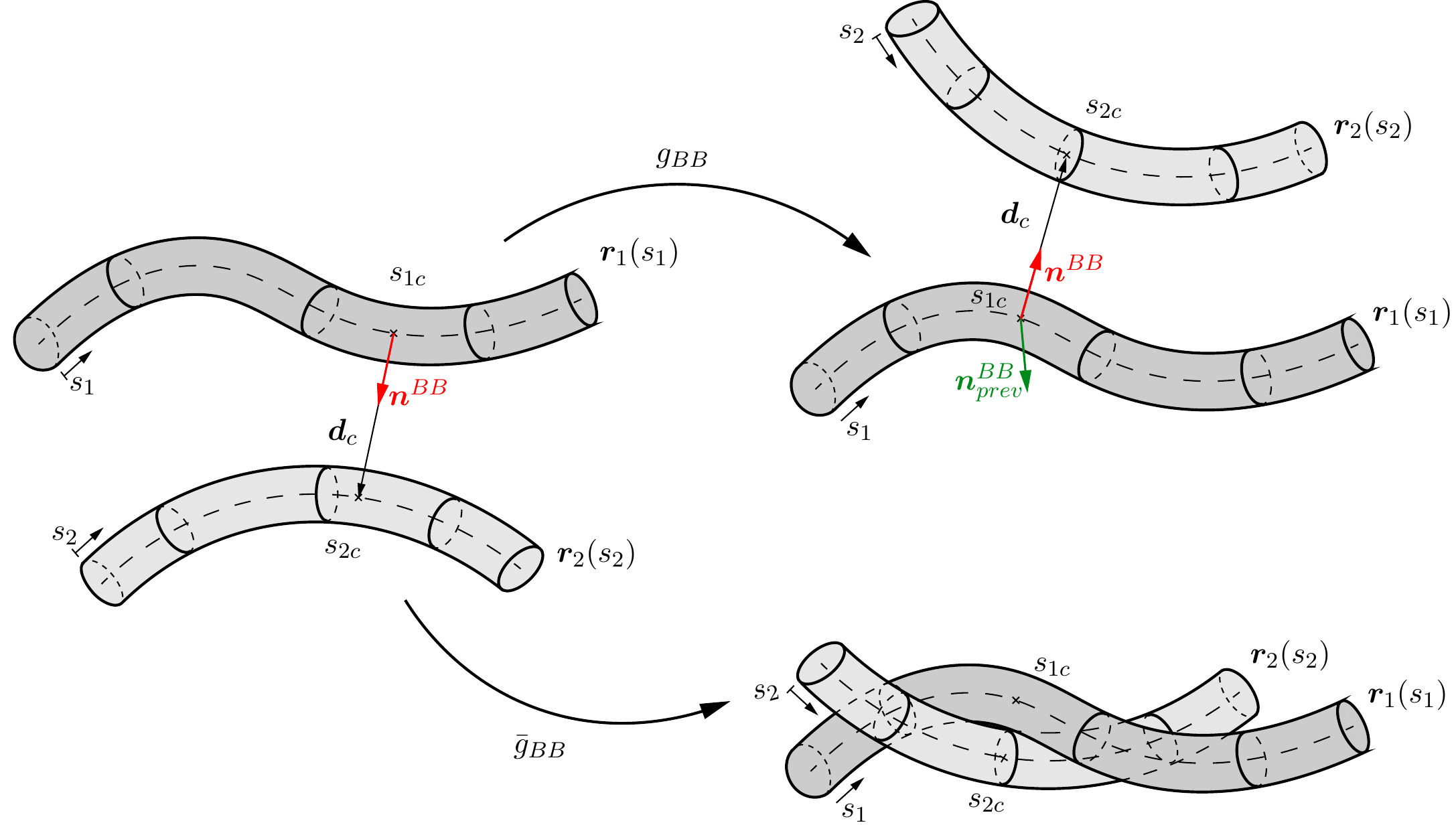}
  \end{minipage}
  \caption{Schematic illustration of a crossing of two beams with
    centerlines~$\beamcenterline_1(\beamarclengthone)$
    and~$\beamcenterline_2(\beamarclengthtwo)$ within one load. Upper branch: with the gap
    function~$\gapbtb$, the crossing remains undetected, since the gap is positive again
    after the crossing, while the current contact normal~$\normalforbtbgap$ (red) has
    reversed its direction compared to the normal~$\historynormalforbtbgap$ of the last
    converged state (green). Lower branch: with the gap function~$\enhancedgap$, the
    reversed normal yields a negative gap, so the crossed configuration is treated as
    penetration and penalized by the contact potential.} \label{fig:beamnewgapvisualization}
\end{figure*}

A well-known challenge in the interaction of slender structures, such as beams with
small cross-sectional radii, is that contact may remain undetected within a load step,
potentially leading to undesired beam crossing as visualized
in~\cref{fig:beamnewgapvisualization}. This issue arises because the gap function
between the centerlines of two beams is not uniquely defined. Various algorithmic
strategies, including load-step or time-step adaptation, have been proposed to improve
the robustness of beam-to-beam contact
formulations~\cite{meierUnifiedApproachBeamtobeam2017}.

In this work, the gap function of the contact potential is enhanced by introducing a
history-dependent normal vector~$\historynormalforbtbgap$, which allows detecting
unintended beam crossings. The modified gap function is defined as
\begin{equation}
  \enhancedgap=
  \operatorname{sign}\!\left( \normalforbtbgap \cdot \historynormalforbtbgap \right)\distbtb - \beamcrossectionradius_1 - \beamcrossectionradius_2 \text{.}
\end{equation}
By incorporating the normal vector~$\historynormalforbtbgap$ of the last converged state
along the nonlinear, quasi-static equilibrium path into the gap evaluation, the formulation retains directional consistency during crossing
events. As a consequence, the contact potential increases in the event of beam
crossing, providing a simple yet robust mechanism to prevent such crossings.

\subsection{Kinematically prescribed solid}
In the present study, only fully kinematically prescribed solids are considered, which
are later used to model the \glspl{ac:MC} that deform the \glspl{ac:FD}. Since the
displacement field is prescribed, no nonlinear equilibrium problem is solved in the
solid domain. The solid body is defined in the reference
configuration~$\domainsolidinitial \subset \mathbb{R}^3$ with
boundary~$\boundarysolid$. The current position of a material point is defined by
\begin{equation}
  \xcurr(\xref) = \xref + \us(\xref)
  \quad\text{and}\quad
  \xref=\transpose{\klr{\xrefX,\xrefY,\xrefZ}},
\end{equation}
where~$\us$ denotes the prescribed displacement field.

\subsection{Beam-to-surface contact}\label{sec:beam-to-surface-contact}
\begin{figure}
  \centering
  \begin{minipage}[b]{\linewidth}
    \centering
    \includegraphics[width=0.8\linewidth]{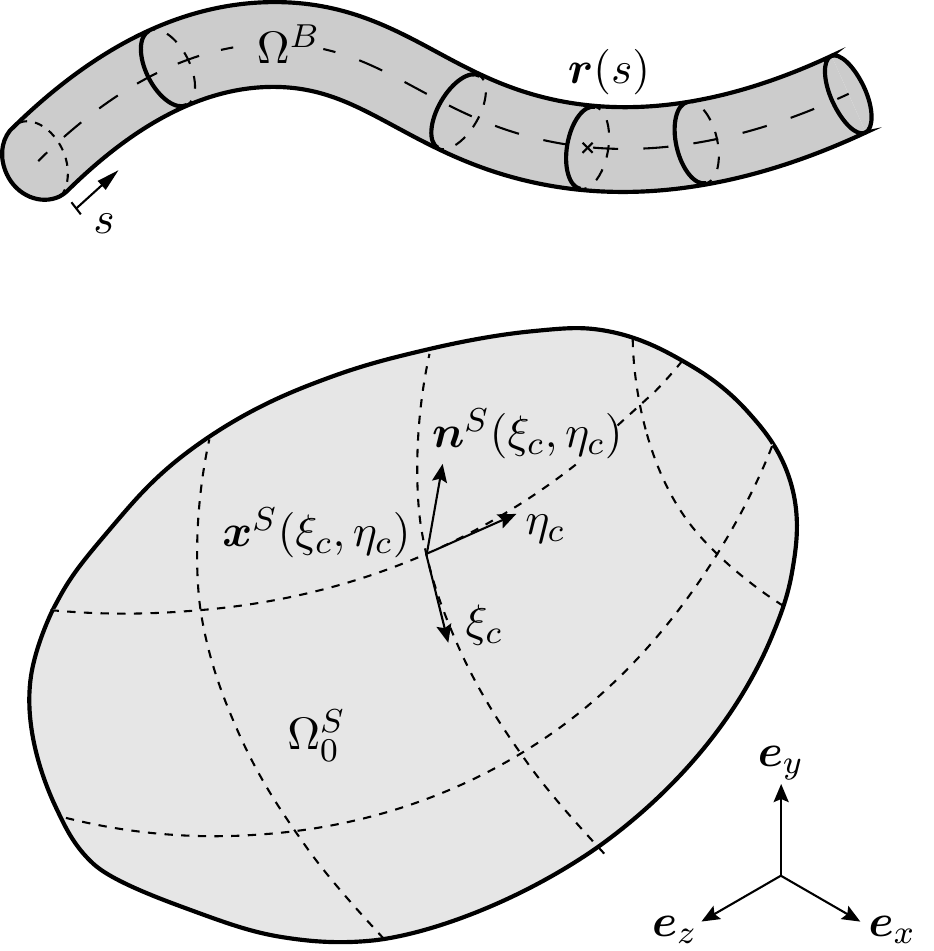}
  \end{minipage}
  \caption{Visualization of the beam-to-surface contact scenario.}
  \label{fig:beamsolidcontact}
\end{figure}
To account for the interaction arising from contact between the one-dimensional beams
and the surface of the solid body, a frictionless contact formulation is considered as
displayed in~\cref{fig:beamsolidcontact}. Therefore, a formulation is considered, where
the contact interaction is described exclusively by the positional field of the beam
centerline~$\beamcenterline(\beamarclength)$, neglecting rotational contact
constraints. The contact conditions are enforced for each point along the beam
centerline. The corresponding closest point on the solid surface is obtained by solving
a unilateral minimal distance problem
\begin{equation}
  \distbtsunilateral(\beamarclength)
  =
  \min_{\solidsurfxi,\solidsurfeta}
  \,
  \distbts\!\left(\beamarclength,\solidsurfxi,\solidsurfeta\right)\text{.}
\end{equation}
The distance function~$\distbts\!\left(\beamarclength,\solidsurfxi,\solidsurfeta\right)$ is given as
\begin{equation}
  \distbts\!\left(\beamarclength,\solidsurfxi,\solidsurfeta\right)
  =
  \norm{\beamcenterline(\beamarclength) - \xcurr^{\solid}(\solidsurfxi,\solidsurfeta)}\text{,}
\end{equation}
where~$\solidsurfxi$ and~$\solidsurfeta$ denote the surface parameters of the solid.
The solution~$(\solidsurfxicp,\solidsurfetacp)$ of the minimization problem implies that the relative
vector between the beam centerline point and the surface point is aligned with
the outward surface normal~$\solidsurfnormal(\solidsurfxicp,\solidsurfetacp)$, \ie~
\begin{equation}
  \beamcenterline(\beamarclength) - \xcurr^{\solid}(\solidsurfxicp,\solidsurfetacp)
  =
  \distbtsunilateral(\beamarclength)\,\solidsurfnormal(\solidsurfxicp,\solidsurfetacp) \text{.}
\end{equation}
The gap function~$\gap(\beamarclength)$ and non-penetration condition can be
stated as
\begin{equation}
  \gap(\beamarclength) = \distbtsunilateral(\beamarclength) - \beamcrossectionradius  \quad \text{~with~}
  \quad
  \gap(\beamarclength)\ge 0 \ \ \forall\,\beamarclength \text{,}
\end{equation}
where~$\beamcrossectionradius$ denotes the cross-section radius of the beam.
The penalty contact potential with the penalty parameter~$\btspenalty$
is defined as
\begin{equation}
  \btspotential
  =
  \frac{1}{2}\,\int_{\beamdomain} \btspenalty\,
  \macaulaybrack{\gap(\beamarclength)}^2 \,\mathrm{d}\beamarclength  \text{.}
\end{equation}
The beam-to-surface contribution to the principle of virtual work follows from
variation of the penalty potential
\begin{equation}
  \virtualbeamtosolidpotential
  =
  \virtualworkbtscontact
  =
  \int_{\beamdomain}\btspenalty
  \macaulaybrack{\gap(\beamarclength)}\,\virtual \gap(\beamarclength)\,
  \mathrm{d}\beamarclength .
\end{equation}
\subsection{Positional coupling as kinematic constraints}\label{sec:positional-coupling}
Kinematic constraints are essential to model complex beam structures such as
\glspl{ac:FD}. Here, a penalty approach is used to suppress the relative motion between
two beam nodes. This constraint, known as a \textit{spherical joint} in the literature,
is referred to as \textit{positional coupling}. In general, arbitrary points of
different beams, typically belonging to different bodies, can be coupled. The coupled
node pairs~$(i,j)$, with nodes~$i$ and~$j$ located on different beams, are collected in
the coupling set~$\couplingset$, and the relative distance vector of a pair reads
\begin{equation}
  \couplingdistance= \beamcenterline_i - \beamcenterline_j \text{,}
\end{equation}
where~$\beamcenterline_i$ denotes the centerline position at node~$i$.
With the scalar penalty parameter~$\pointcouplingparam > 0$, the coupling potential is
\begin{equation}
  \couplingpotential
  =
  \sum_{(i,j)~\in~\couplingset}
  \frac{1}{2}\,\pointcouplingparam\,
  \norm{\couplingdistance}^2 \text{.}
\end{equation}
Since coupled nodes coincide in the reference configuration, it penalizes any relative
displacement between them. Its first variation contributes to the principle of virtual
work as
\begin{equation}
  \virtual \couplingpotential =
  \virtualworkcoupling
  =
  \sum_{(i,j)~\in~\couplingset}
  \pointcouplingparam\,
  \couplingdistance\cdot
  (\virtualbeamcenterline_i-\virtualbeamcenterline_j) \text{.}
\end{equation}
Further details regarding formulation and implementation
can be found in~\cite{steinbrecherVariationallyConsistentBeamtobeam2026}.
\subsection{Local coordinate transformation}\label{sec:coordinate-transform}
For actual \glspl{ac:FD} geometries, it is not possible to prescribe all boundary
conditions solely with respect to a single global Cartesian coordinate system. In
particular, constraints on axial or radial motion are more conveniently defined with
respect to a local coordinate system that is aligned with the wire orientations of a
\gls{ac:FD}. To this end, a local orthonormal basis
$\{\rotunitlocx,\rotunitlocy,\rotunitlocz\}$ is introduced at selected points of the
domain and a corresponding rotation matrix $\rotglmatrix \in \mathrm{SO}(3)$ is
obtained by arranging the basis vectors as rows,
\begin{equation}
  \rotglmatrix =
  \begin{bmatrix}
    \transpose{\rotunitlocx} \\
    \transpose{\rotunitlocy} \\
    \transpose{\rotunitlocz}
  \end{bmatrix}, \quad \text{ with }
  \quad
  \transpose{\rotglmatrix}\rotglmatrix=\identitymatrix \text{.}
\end{equation}
A displacement vector expressed in global coordinates,
$\displacement^{\rotglob}$, is transformed to the local coordinate system according
to
\begin{equation}
  \displacement^{\rotloc}
  =
  \rotglmatrix\,
  \displacement^{\rotglob}.
\end{equation}

\section{Structural mechanics model for interacting braided neurovascular stents}\label{sec:FD-model}
The formulations from~\cref{sec:modeling} are integrated into a mechanical model of
a~\gls{ac:FD}, and its contact interaction with a simplified \gls{ac:MC} model is
described. The choice of each formulation is motivated by the mechanical behavior of
the~\gls{ac:FD} that it has to capture, and all formulations are coupled within a
consistent model in~\fourc~\cite{4C}.
%

\subsection{Mechanical modeling of flow diverters based on interacting beams}\label{sec:FD-model-explanation}
The thin wires for the \gls{ac:FD} are modeled using the introduced geometrically exact
Simo--Reissner beam theory described in~\cref{sec:beamtheory}. This beam formulation is
particularly suited for the modeling of the wires, since this beam formulation is able
to account for large rotations which are expected during the deployment of
a~\gls{ac:FD}. By modeling the beam as a one-dimensional Cosserat continuum with
independent translational and rotational degrees of freedom, it enables a consistent
description of axial, bending, torsional, and shear deformation modes. In addition, an
initially curved, stress-free reference configuration can be naturally incorporated
into the present beam formulation. This enables the spatially varying curvature of the
interwoven~\gls{ac:FD} structure to be accounted for directly, without the need to
first determine the corresponding curved equilibrium state, as is often required in
beam formulations based on an initially straight reference axis. Furthermore, since the
actual stress-free reference configuration of the \gls{ac:FD} is not known, possible
residual stresses associated with the manufacturing or braiding process are neglected.
The generated interwoven geometry in~\cref{sec:fdparametric} is therefore assumed to
represent the stress-free reference configuration throughout this work.

From a modeling perspective it is essential to capture the contact interaction between
the individual wires, since modeling strategies based on kinematic constraints, {\eg}
hinges or joints, may lead to non-realistic behavior. Therefore, the selected contact
formulation is suitable to account for the interactions between the individual fibers
as long as the contact angle remains large.

At each device end, every wire of one braiding direction meets exactly one wire of the
opposite direction at a common end point. The two end nodes form a \textit{coincident
end-node pair}, whose positions are coupled as described
in~\cref{sec:positional-coupling}, while their rotations remain independent. This
prevents fraying of the wire ends. In contrast to closed-end designs, in which each
wire continues smoothly into the opposite braiding direction, the rotations of the
wires are not continuous across the coincident end-node pairs, and the modeled device
therefore behaves as an open-end stent. The influence of open and closed ends on the
mechanical behavior of braided stents is investigated
in~\cite{shanahanLoopedEndsOpen2017,ubachsComputationalModelingBraided2023}.
The resulting overall virtual work for the \gls{ac:FD} model can be stated as:
\begin{equation}
  \virtualworkbeamtot
  =
  \virtualworkbeamint
  +
  \virtualworkbtbcontact
  -
  \virtualworkbeamext
  +
  \virtualworkcoupling \text{.}
\end{equation}
The wires are discretized with geometrically exact Simo--Reissner beam finite
elements~\cite{meierGeometricallyExactBeam2018}. The beam centerline is interpolated with
cubic Hermite polynomials based on the nodal positions and tangents at the two element end
nodes. This yields a smooth, $C^1$-continuous centerline, which is advantageous for
beam-to-beam contact~\cite{meierUnifiedApproachBeamtobeam2017}. The cross-section
orientation is interpolated by an objective triad
interpolation~\cite{crisfieldObjectivityStrainMeasures1999} based on the nodal rotations
of the two element end nodes and an additional element midpoint node.
Further details regarding the beam discretization can be obtained
from~\cite{meierGeometricallyExactBeam2018,meierUnifiedApproachBeamtobeam2017}.
The resulting nonlinear
problem is solved incrementally using Newton's method~\cite{wriggersNonlinearFiniteElement2008},
where each load step increases the load parameter
from~$\loadstepparameter_n$ to~$\loadstepparameter_{n+1}$, with~$\loadstepparameter\in[0,1]$.
Another important aspect concerns the selected
material model. As introduced in~\cref{sec:beamtheory}, only a linear elastic
constitutive model is considered for the individual wires of the \gls{ac:FD}. From a
modeling perspective, a superelastic material model accounting for the austenitic and
martensitic phases of
Nitinol~\cite{auricchioShapememoryAlloysMacromodelling1997,lagoudasUnifiedThermodynamicConstitutive1996}
may be better and provide a more accurate description of the mechanical
response~\cite{mckennaExperimentalEvaluationMechanics2020,kimMechanicalModelingSelfexpandable2008},
depending on the load scenario. However, within other biomedical application
cases~\cite{Perrin2016,Hemmler2019} the Nitinol material was successfully approximated
by linear elasticity as long as the strains remain sufficiently small and nonlinear
material effects due to thermal changes are neglected. Based on these assumptions, the
present approach adopts a linear elastic model under the condition that the strain
levels remain within the regime where superelastic phase transformation is not
activated. To ensure the validity of this assumption, the maximum strain within the
wires is evaluated and compared against the characteristic transformation strain of
superelastic Nitinol. Hence, if the computed strains remain below this threshold, the
assumption of the material being linear elastic is reasonable.

\subsubsection{Geometric description of an interwoven \glsentryname{ac:FD}}\label{sec:fdparametric}
\begin{figure*}[t]
  \centering
  \begin{minipage}[b]{0.8\textwidth}
    \includegraphics[width=\linewidth]{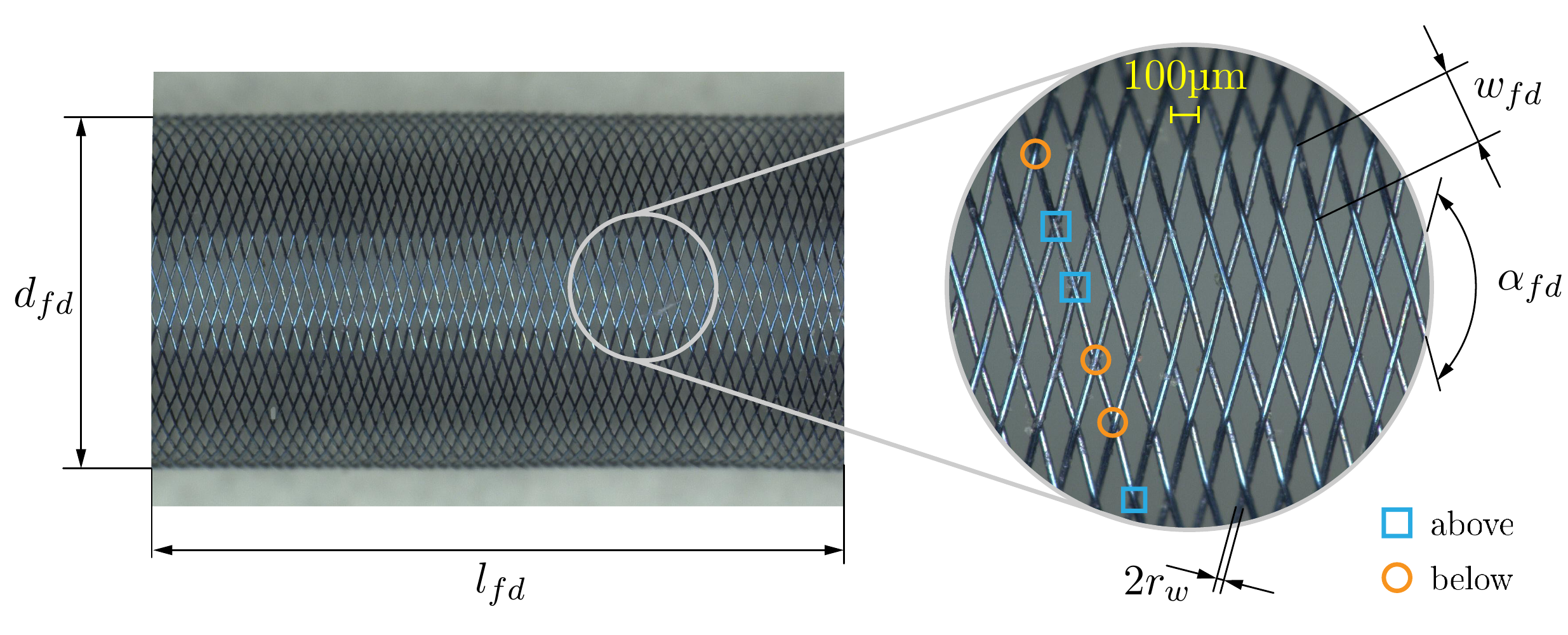}
  \end{minipage}
  \caption{The visualization shows a small segment of a real~\gls{ac:FD}, including a detailed
    view of the braiding pattern and the key dimensions. From the zoomed-in view,
    the wire pattern can be identified. The orange circle (below)
    and the blue box(above), indicates if the considered wire passes above the
    wires along the other braiding direction. The wire radius~$\rfdwire$,
    braiding angle~$\fdbraidingangle$, and the distance~$\fdw$ between two individual
    wires are shown together with a scale bar of~$\qty{100}{\micro\meter}$.
    The length~$\lfd$ and diameter~$\dfd$ are included for completeness and are
    visualized on the smaller segment.\label{fig:realfd} }
\end{figure*}

\begin{figure*}
  \centering
  \begin{minipage}[b]{\textwidth}
    \centering
    \includegraphics[width=\linewidth]{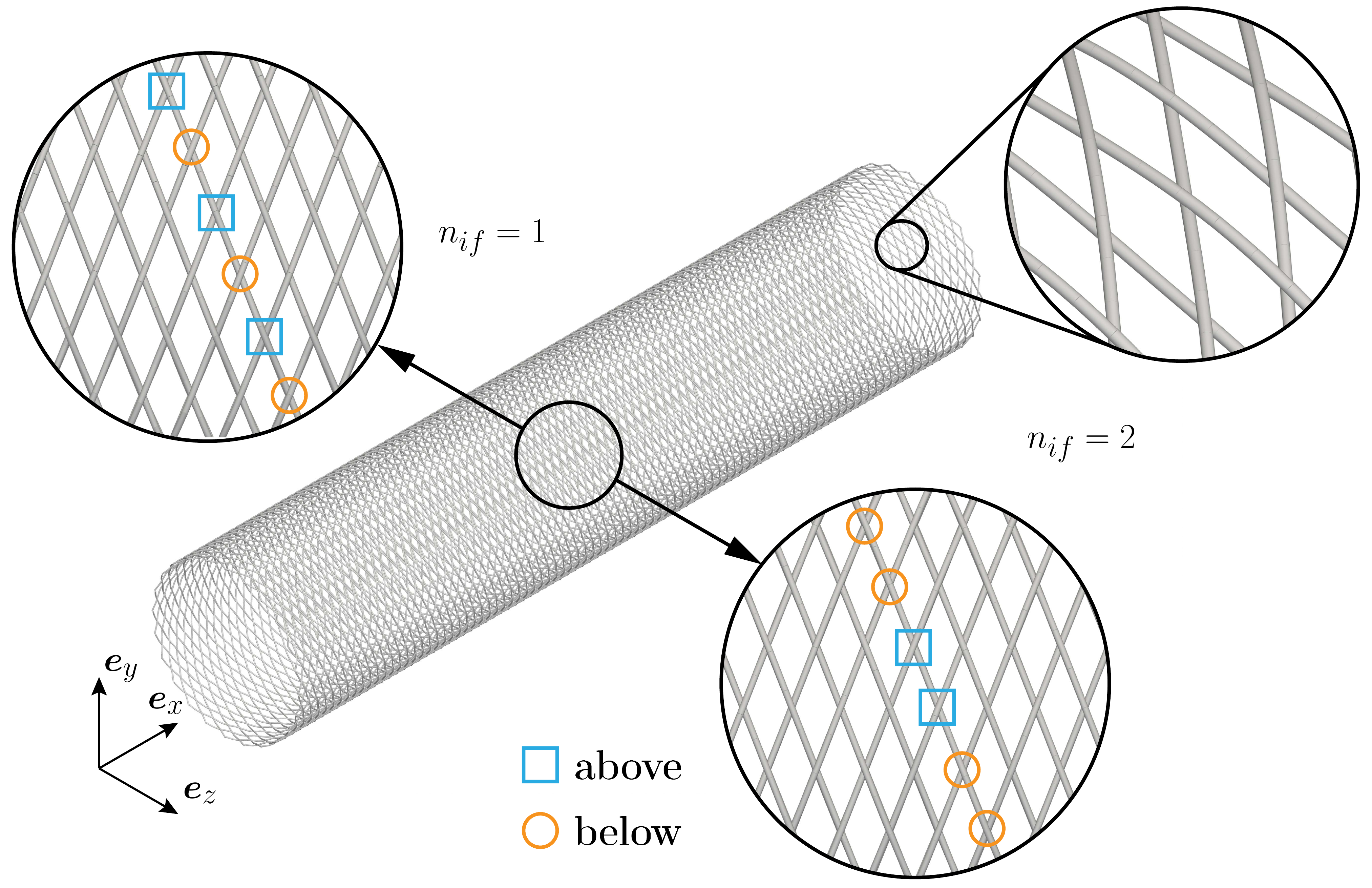}
  \end{minipage}
  \caption{Visualization of an interwoven~\gls{ac:FD} generated from the presented parametric equations, including a detail view highlighting the crossing wires toward the end of the~\gls{ac:FD}.
    The influence of the parameter~$\fdinterwoovenfamiliy$ is illustrated in a detailed view for the values~$1$
    and~$2$. The resulting braiding pattern is indicated at each crossing (or at
    every second crossing for a single wire), where circles and boxes denote
    whether the wire passes above or below its neighboring wire with other braiding directions.
  }
  \label{fig:meshfd}
\end{figure*}

A~\gls{ac:FD} can be represented as a cylindrical structure composed of~$24$ to~$64$
total helical wires~$\nwire$ that are arranged in two counter-wound
directions. At each overlap the prescribed braiding angle~$\fdbraidingangle$ can be measured,
as indicated in the snippet of a real \gls{ac:FD} in~\cref{fig:realfd}.
Additionally, a radial interweaving or braiding pattern
can be identified from~\cref{fig:realfd} in each crossing, avoiding geometric intersections between
two counter-wound wires. Therefore, a braided \gls{ac:FD} is typically characterized by its geometric parameters,
which include the length~$\lfd$, the mean cylinder radius~$\rfd$, the braiding
angle~$\fdbraidingangle$ and the radius~$\rfdwire$ of the individual wires.
Some manufacturers additionally incorporate a few
wires with different material properties and radii~$\rfdwire$ to enhance radiopacity
and improve visibility under medical imaging. The following approach is restricted to
wires with identical circular cross-sections of radius~$\rfdwire$ and does
not account for a variation within the wire diameters.
Therefore, the centerline of an individual wire is described by a
smooth parametric curve~\( \fdxparametricsingle \in \mathbb{R}^3 \)
which is parameterized by the dimensionless coordinate~\( \parametert \in [0,1] \).
The parameter~\(\fdbradingdirection \in \{-1,+1\}\) ensures that each wire belongs
to one of the two families according to the clock or counter-clockwise orientation
of the wire.
The complete parametric representation of a fully braided \gls{ac:FD} is obtained
by~$\fdxparametric$, which considers all individual wires~\( \fdwireindex = 0,\dots,{\nwire-1} \).
For a fixed pair~\( (\fdbradingdirection,\fdwireindex) \), the wire centerline is given by the parametric
description

\begin{equation}
  \label{eq:fd_centerline}
  \fdxparametricsingle
  =
  \begin{bmatrix}
    \parametert\, \fdtotalhelixlength \cos\!\left(\fdhelixangle\right)
    \\[4pt]
    \fdampt\,
    \cos\!\left(\fdphaseanglefct\right)
    \\[4pt]
    \fdampt\,
    \sin\!\left(\fdphaseanglefct\right)
  \end{bmatrix}.
\end{equation}
The arc length of a rotated wire~$\fdtotalhelixlength =\lfd / \cos\!\left(\fdhelixangle\right)$
is  given with respect to the helix angle~\(\fdhelixangle=\frac{\fdbraidingangle}{2}\). The circumferential phase angle~\( \fdphaseanglefct \) is defined by
\begin{equation}
  \label{eq:fd_theta}
  \fdphaseanglefct
  =
  \fdbradingdirection
  \left(
  \frac{2\pi \fdwireindex}{\nwire}
  -
  \frac{\parametert \fdtotalhelixlength}{\rfd}
  \sin\!\left(\fdhelixangle\right)
  \right) \text{,}
\end{equation}
accounting for the helix structure.
The radial interweaving between wire families is obtained by a direction-dependent
radius modulation, where its amplitude is controlled by the parameter~\( \fdinterwoovenamplitude \).
The radial coordinate~$\fdampt$ along the helix is varied such that crossing
wires can be arranged above or below each other:
\begin{equation}
  \label{eq:fd_radius}
  \fdampt
  =
  \rfd + \fdbradingdirection \fdinterwoovenamplitude
  \sin\!\left(
  \frac{2\pi}{\fdwavelength}\,\parametert \fdtotalhelixlength + \frac{\pi}{4}
  \right) \text{.}
\end{equation}
Hence, the interwoven pattern accounting for the change in
radial direction is based on the sinusoidal function along the parametric curve.
The wavelength~\( \fdwavelength \) of the interwoven pattern along the helical arc length
can be obtained via
\begin{equation}
  \label{eq:fd_pattern_period}
  \fdwavelength
  =
  \frac{2\pi\,\rfd\,\fdinterwoovenfamiliy}{\sin\!\left(\fdhelixangle\right)\nwire} \text{.}
\end{equation}
Within~\cref{eq:fd_pattern_period} the variable~$\fdinterwoovenfamiliy$ is used to
determine if braiding occurs at every intersection or only at each
$\fdinterwoovenfamiliy$-th intersection of the wires and therefore controls the
resulting braiding pattern as visualized in~\cref{fig:meshfd}. Note that the respective
sine function from~\cref{eq:fd_radius} must be chosen with an appropriate
amplitude~$\fdinterwoovenamplitude$ to avoid intersections. Generally, the radius of the
wire is a recommended starting point for the amplitude. For higher numbers of~$\fdinterwoovenfamiliy$ it might
be necessary to enhance the sine function to avoid overlapping within the initial
geometry. By simply increasing the radial amplitude parameter~\( \fdinterwoovenamplitude \)
the initial geometry of the \gls{ac:FD} may not be represented accurately.
The parametric equations can be efficiently implemented within the open-source
finite-element beam generator~\beamme~\cite{BeamMe} to create suitable meshes
for beam finite elements.

\subsubsection{Local coordinate system for boundary conditions at end points}
\begin{figure}
  \centering
  \begin{minipage}[b]{\linewidth}
    \centering
    \includegraphics[width=0.8\linewidth]{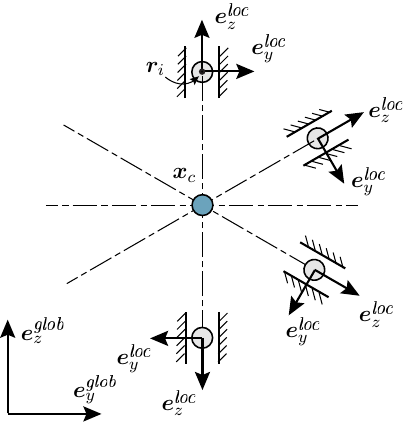}
  \end{minipage}
  \caption{Visualization of the transformed local coordinate system in combination with
    boundary conditions. Here, the end nodes are constrained along the local~$\rotunitlocy$
    resulting in radial displacements towards the center~$\rotcenter$
    of the flow diverter.}
  \label{fig:locsys}
\end{figure}

At the end points of each wire of a \gls{ac:FD}, a locally rotated coordinate system is
introduced to allow a realistic radial displacement of the structure while restricting
rigid body motions or global rotation modes. This local coordinate system is defined
individually for each endpoint of a wire. The local unit vector~$\rotunitlocz$ is
chosen to align with the radial direction pointing from the center~$\rotcenter$ to node
position~$\rotnodeposition$
\begin{equation}
  \rotunitlocz
  =
  \frac{\rotnodeposition - \rotcenter}
  {\norm{\rotnodeposition - \rotcenter}}.
\end{equation}
The center~$\rotcenter$ is the orthogonal projection of the node onto the centerline of
the \gls{ac:FD}.
The direction~$\rotunitlocx$ is chosen to be aligned with the global
axial direction, \ie~$\rotunitlocx = \rotunitglobx$. Hence,~$\rotunitlocz$
and~$\rotunitlocx$ are orthogonal unit vectors, and the orthonormal basis is completed by
\begin{equation}
  \rotunitlocy
  =
  \rotunitlocz \times \rotunitlocx \text{.}
\end{equation}
Any applied boundary conditions at the ends are subsequently
prescribed with respect to this local coordinate system, as displayed in~\cref{fig:locsys}.

\subsection{Interaction between \glsentryname{ac:FD} and \acrlong{ac:MC}}\label{sec:Interaction-explanation} The introduced beam-to-surface
contact approach is employed to capture the interaction between a surface and the
individual wires of the device in a physically realistic manner. Hence, the approach
permits the device to adapt naturally to the surrounding geometry during insertion and
expansion, thereby enabling a realistic simulation of the mechanical interaction
processes occurring during deployment. In contrast to other approaches, such as
kinematic multi-point constraint formulations, this contact-based approach allows for
relative sliding and separation between the wires and the surface, which is essential
for accurately representing the deployment and can be used in future cases for
placement of the device. From a numerical perspective, the presented approach provides
a robust and flexible methodology for solving the resulting mixed-dimensional problem
in a stable manner.

In the following section, the three-dimensional solid represented by a \gls{ac:MC} or a
crimping device is, within the scope of this work, simplified to a rigid surface and
displayed with uniform wall thickness. This simplification can be employed, since the
internal stress states of the solid are not within the focus of this work and the
displacement field must be prescribed on the contact side due to the considered
examples. From a geometric point of view, the tube is characterized by an inner
diameter~$\dmcatinner$ and an outer diameter~$\dmcatouter$, which are related by a
constant thickness~$\mcatthickness$. Furthermore, this surface is considered to be a
sufficiently smooth boundary representation to prescribe the displacement field.
The resulting total virtual work~$\virtualworktot$ of the coupled system accounting for
all interactions is defined as
\begin{equation}
  \virtualworktot
  =
  \virtualworkbeamtot
  +
  \virtualworkbtscontact
  \text{.}
  \label{eq:totoalvirtualwork}
\end{equation}
The further discretizations of the presented virtual works can be found
within~\cite{ivodissertion}.

\section{Numerical examples}\label{sec:numerical-examples}
In the following, a series of benchmark or validation cases is presented, to verify the
modeling assumptions and investigate the correctness and predictability of the modeling
approach. The first two cases are taken
from~\cite{jedwabStudyGeometricalMechanical1993}, which have also been used for
comparisons in~\cite{shanahanLoopedEndsOpen2017,maComputerModelingDeployment2012}. The
third example is taken from~\cite{shapiroVariablePorosityPipeline2014} and outlines the
relevant metrics. It was also studied by fast virtual stenting
approaches~\cite{jeken-ricoVirtualFlowDiverter2024}.

\subsection{Tension test case}\label{sec:tensile-example}
This tensile test is conducted as a literature-based benchmark to assess whether the
proposed model reproduces the characteristic coupling between axial elongation, radial
contraction, and axial force of a \gls{ac:FD} under quasi-static loading conditions.
This loading regime is of particular clinical and mechanical relevance, since the
resulting diameter--length relationship represents a key criterion in the
preinterventional selection of an appropriate \gls{ac:FD}, while comparable loading
conditions also arise during deployment. The present example should therefore not be
regarded as a direct reproduction of the experiment, where the reference data were
obtained for a \textit{Wallstent}~\cite{jedwabStudyGeometricalMechanical1993}. Instead,
the present model follows the proposed \gls{ac:FD} approach, such that some geometric
quantities and modeling assumptions cannot be matched exactly. Moreover, the analytical
reference solution is based on idealized open-coiled helical spring models with
rotationally constrained ends and neglects interactions between individual wires.
Accordingly, the comparison is mainly interpreted in terms of trends, order of
magnitude, and the consistency of the predicted deformation and force response.

\begin{figure}
  \centering
  \begin{minipage}[b]{\linewidth}
    \centering
    \includegraphics[width=0.95\linewidth]{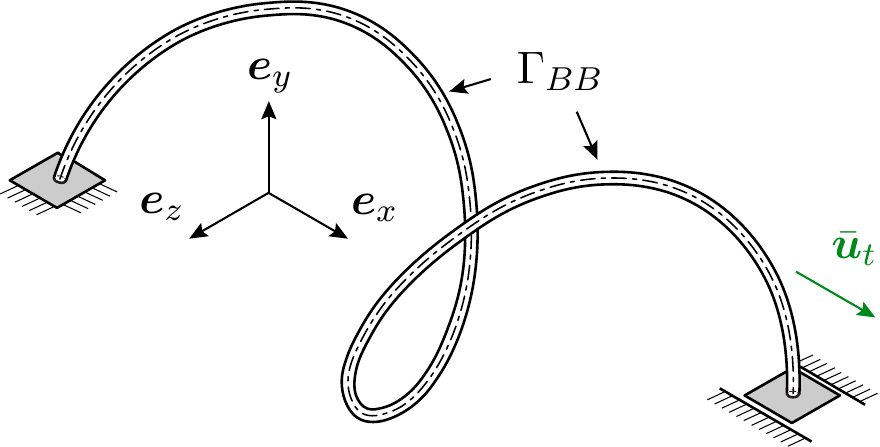}
  \end{minipage}
  \caption{Visualization of the boundary conditions for a single helical wire within a uniaxial loading scenario.}
  \label{fig:tensile-test-overwiew}
\end{figure}

\begin{figure}
  \centering
  \begin{tikzpicture}
    \node[inner sep=0pt] (img) {\includegraphics[width=\dimexpr\linewidth-2em\relax,
        trim={200 139 160 170},clip]{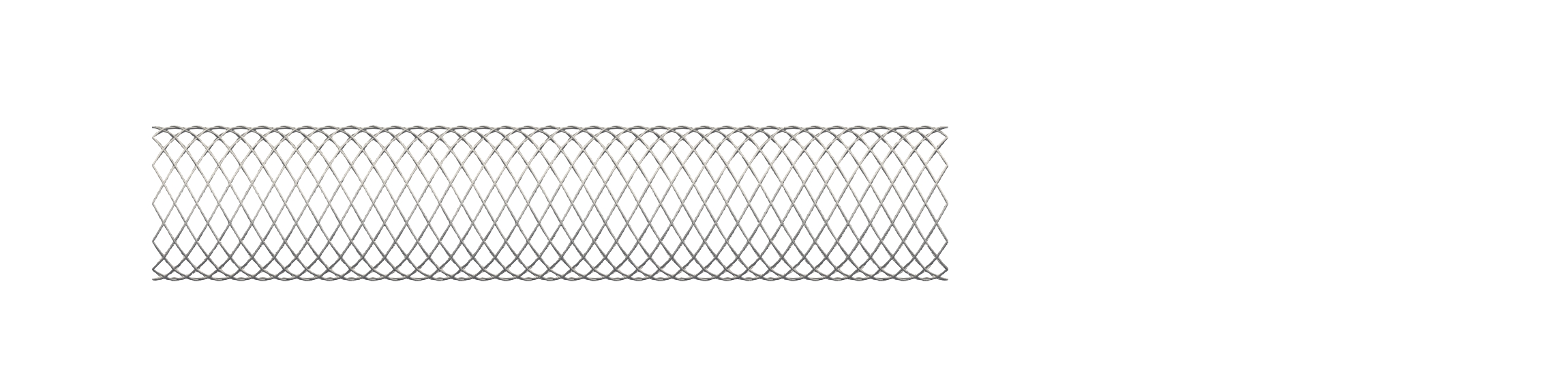}};
    \node[anchor=east] at (img.west) {(a)};
    \node[anchor=east,inner sep=0pt] at ([xshift=-0.5cm,yshift=0.5cm]img.east)
    {\includegraphics[height=1.1cm]{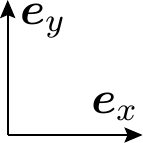}};
  \end{tikzpicture}\\[1.5ex]
  \begin{tikzpicture}
    \node[inner sep=0pt] (img) {\includegraphics[width=\dimexpr\linewidth-2em\relax,
        trim={200 139 160 170},clip]{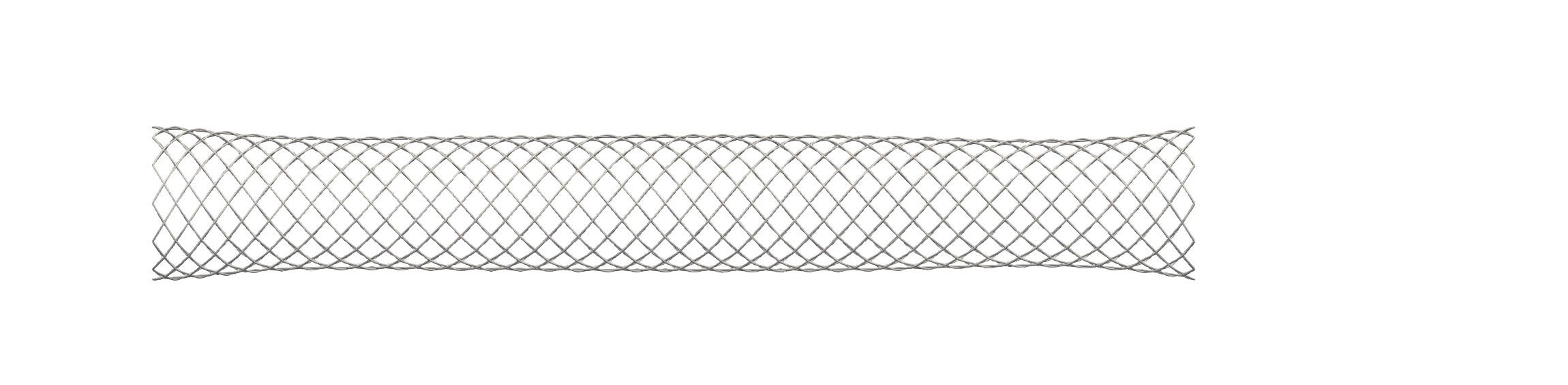}};
    \node[anchor=east] at (img.west) {(b)};
  \end{tikzpicture}\\[1.5ex]
  \begin{tikzpicture}
    \node[inner sep=0pt] (img) {\includegraphics[width=\dimexpr\linewidth-2em\relax,
        trim={200 139 160 170},clip]{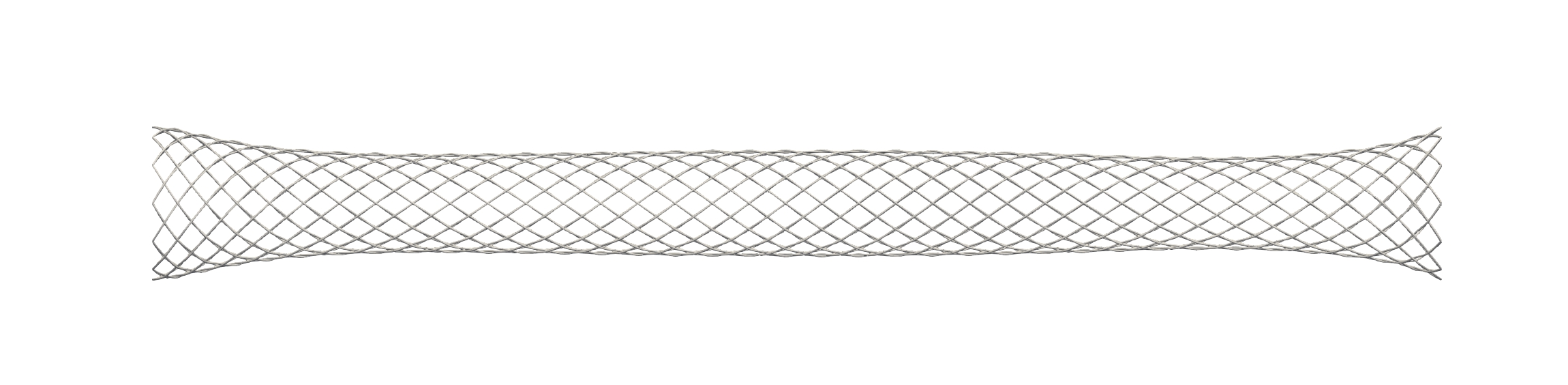}};
    \node[anchor=east] at (img.west) {(c)};
  \end{tikzpicture}

  \caption{Visualization of the deforming stent during the uniaxial tensile test: (a) initial
    configuration, (b) deformed configuration at an intermediate state, and (c) elongated
    device in the final configuration.}
  \label{fig:tensile-test-visualization}
\end{figure}

For the following example, the \gls{ac:FD} is discretized with \num{80} beam elements
per wire resulting in a total number of \num{3864} nodes and \num{23256}
\glspl{ac:DOF}. The parameters used for the benchmark comparison are based on
\cite{jedwabStudyGeometricalMechanical1993} and are summarized, together with selected
simulation parameters, in~\cref{tab:example-01-parameters}.
\begin{table}
  \centering
  \setlength{\tabcolsep}{4pt}
  \begin{tabular}{l c c}
    \hline
    \textbf{Parameter}                                & \textbf{Value} & \textbf{Unit}                 \\
    \hline
    Number of wires~($\nwire$)                        & ~$12$          & ~$\kle{-}~$                   \\
    Number of turns~($\nfdturns$)                     & ~$3.1$         & ~$\kle{-}~$                   \\
    Helix angle~($\fdhelixangle$)                     & ~$61.73$       & ~$\kle{\unit{\degree}}~$      \\
    Wire radius~($\rfdwire$)                          & ~$0.11$        & ~$\kle{\unit{\milli\meter}}~$ \\
    Cylinder radius~($\rfd$)                          & ~$8.355$       & ~$\kle{\unit{\milli\meter}}~$ \\
    Length~($\lfd$)                                   & ~$8.75$        & ~$\kle{\unit{\centi\meter}}~$ \\
    Young’s modulus~($\beamyoungsmod$)                & ~$206$         & ~$\kle{\unit{\giga\pascal}}~$ \\
    Beam point penalty parameter ($\btbpointpenalty$) & ~$500$ & ~$\kle{-}~$ \\
    \hline
  \end{tabular}
  \caption{Parameters for the tensile test case, based
    on~\cite{jedwabStudyGeometricalMechanical1993}.}
  \label{tab:example-01-parameters}
\end{table}
For the boundary conditions of a tensile test, one end is fully clamped
as depicted in~\cref{fig:tensile-test-overwiew}.
At the other end, a displacement~$\displacementtensiletest$
is prescribed in axial direction to elongate the \gls{ac:FD}.
The displacement~$\displacementtensiletest$ is gradually increased until the \gls{ac:FD} reaches its
maximum length of~$\qty{15}{\centi\meter}$.
Along each wire a potential contact boundary~$\btbcouplingcondition$ is considered
to find the contact between individual wire pairs.
In~\cref{fig:tensile-test-visualization} the results are visualized and show that the
device elongates along the axial direction. The average diameter of the \gls{ac:FD}
decreases towards the final end state of the test, but shows a variation depending on
the axial position.

\begin{figure*}
  \centering
  \begin{minipage}[b]{0.48\textwidth}
    \includegraphics[width=\linewidth]{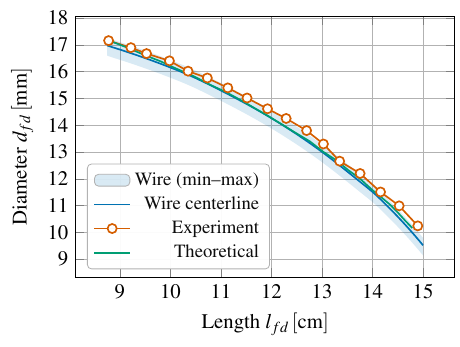}
    \caption{Relation between the diameter~$\dfd$ and length~$\lfd$ of a \gls{ac:FD}
      during an uniaxial tensile test in comparison to the experimental and theoretical results
      from~\cite{jedwabStudyGeometricalMechanical1993}.}
    \label{fig:tensile-test-d-l}
  \end{minipage}
  \hfill
  \begin{minipage}[b]{0.48\textwidth}
    \includegraphics[width=\linewidth]{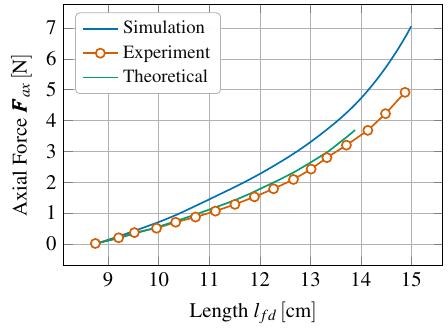}
    \caption{Axial force~$\tensiletestaxialforce$ during an uniaxial tensile test compared to the experimental and theoretical forces
      from~\cite{jedwabStudyGeometricalMechanical1993}.\\}
    \label{fig:tensile-test-f-l}
  \end{minipage}

\end{figure*}
In the following, the results are discussed regarding three principal criteria. First,
the strain measures are investigated to assess the validity of the material
assumptions. Second, the global mechanical response of the device is analyzed in terms
of the diameter--length relationship and the predicted axial force--length
relationship. Third, these results are interpreted and compared to the corresponding
theoretical and experimental results reported by the literature. For each of these
evaluation steps, the relevant post-processing procedures are briefly outlined.


To justify the use of a linear-elastic constitutive law for the beam model in the
present example, the resulting strains are discussed and related to a general
three-dimensional continuum. For the Simo--Reissner beam theory the material
deformation measures~$\beammatdefmeas$ and~$\beammatcurvevector$ represent axial/shear
deformation and torsion/bending, respectively. In particular, the first component of
$\beammatdefmeas$ corresponds to axial strain, while the second and third components of
$\beammatcurvevector$ are associated with bending curvature. According to the relation
between the one-dimensional beam measures and the associated three-dimensional
continuum strain field, the axial Green--Lagrange strain varies linearly over the beam
cross-section and for a circular cross-section of a beam. By neglecting torsion and
shear deformation, a conservative estimate of the maximum outer-fiber axial strain can
be assumed as:
\begin{equation}
  \strainmax
  \approx
  \abs{\beammatdefmeas_1}
  +
  \beamcrossectionradius
  \sqrt{
    \beammatcurvevector_2^2+\beammatcurvevector_3^2
  }\label{eq:strain-comparison-measure}.
\end{equation}
In the present example, this post-processing strain is evaluated as a screening measure for
the onset of material nonlinearity. Hence, the maximum value of this measure is found
to be~$\strainmax= \num{8.7383e-4}$. Since this value
remains comparably small, no pronounced superelastic material response is expected
and the use of a linear-elastic constitutive description for
the beam model is considered as justified for this example.

In~\cref{fig:tensile-test-d-l} the change within the diameter along the device length
during the tensile test compared to the experimental and theoretical results are
presented. Since the considered modeling approach accounts for interwoven wires, the
diameter may change according to the braiding pattern. Considering only a single wire
to estimate the diameter as reference quantity is considered not to be sufficient.
Therefore, all nodes with an axial component initially located between
$\qty{42.75}{\milli\meter}$ and~$\qty{43.75}{\milli\meter}$ are considered for the
computation of the average diameter. Due to the symmetric results, it is sufficient to
consider only a quadrant within the radial plane of the tensile test for computation.
The associated minimum and maximum values define the diameter envelope within the
evaluation length visualized as a range in~\cref{fig:tensile-test-d-l} and indicate the
variations within the minimal inner and maximal outer diameter due to the braiding of
the \gls{ac:FD}.

By inspecting~\cref{fig:tensile-test-d-l}, it can be seen that the numerical
predictions follow the experimental and theoretical trends closely over the loading
range, reproducing the characteristic monotonic diameter reduction during axial
elongation.

In~\cref{fig:tensile-test-f-l}, the axial force during the uniaxial tensile test is
compared to the theoretical and experimental values from the literature. For the
presented modeling approach, the total reaction force~$\tensiletestaxialforce$ in axial
direction is calculated by summation of the axial forces of each individual wire at the
prescribed boundary at the top of the \gls{ac:FD}. The axial force increases with
elongation of the \gls{ac:FD}, and the model captures the expected stiffening trend.
However, an overprediction relative to the reference data is observed at larger
elongations.

Overall, the model qualitatively reproduces the characteristic coupled response of the
braided \gls{ac:FD}, with axial elongation leading to radial contraction and an
increasing axial reaction force. Quantitative agreement with the reference data,
however, deteriorates at larger elongations, particularly with respect to the axial
force. The observed discrepancies may arise from differences between the modeled and
experimental geometries or from simplifications introduced by the modeling assumptions.
Further sensitivity analyses and parameter calibration, complemented by validation
against additional experimental data, could provide additional insight into the
principal sources of discrepancy and help improve the predictive capability of the
model.

\subsection{Compression test case}\label{sec:crimping-test}
\begin{figure*}
  \centering
  \begin{minipage}[b]{\textwidth}
    \centering
    \includegraphics[width=\linewidth]{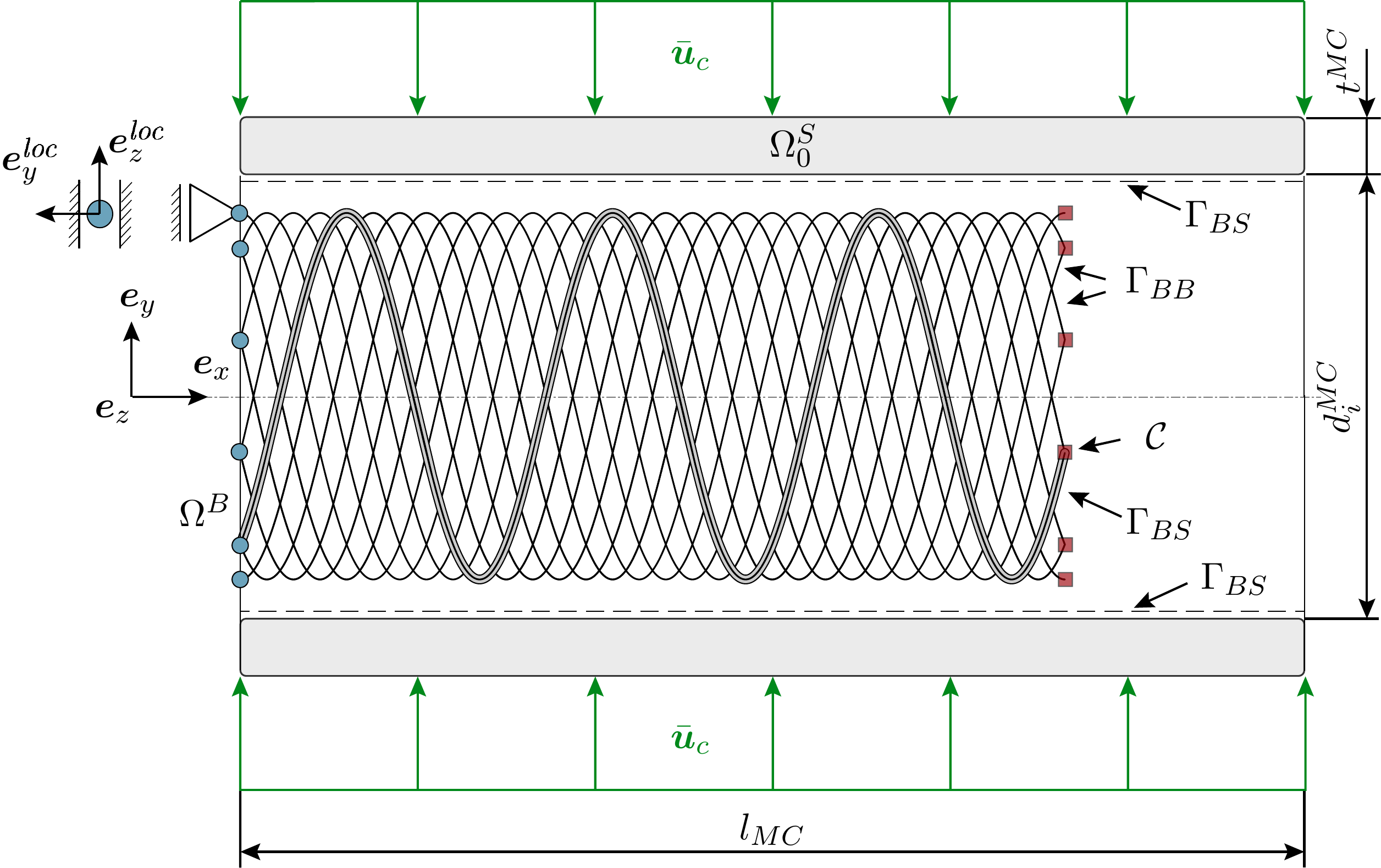}
  \end{minipage}
  \caption{Schematic drawing of the crimping scenario of a \gls{ac:FD}
    subjected to a compressing tube, including the boundary conditions. Coincident
    end-node pairs subject to positional coupling are marked in blue (left end) and
    red (right end). The inset shows the local coordinate system at the left end, viewed
    along the device axis.}
  \label{fig:compressesion-test-overview}
\end{figure*}

To verify that the acting pressure after the deployment of a \gls{ac:FD} is comparable
to established approaches, a simplified compression test case is considered. The
objective of this test is to ensure that the pressure from the deformed~\gls{ac:FD}
acting on a potential artery remains within a physically meaningful range. The
resulting forces have received increasing attention, since they characterize the
mechanical interaction between a self-expanding stent and the surrounding vessel after
deployment. In particular, the remaining opening force during self-expansion is
commonly referred to as the chronic outward force~\cite{CABRERA2017252}. The following
numerical example is based on a similar setup proposed
in~\cite{jedwabStudyGeometricalMechanical1993} and is illustrated
in~\cref{fig:compressesion-test-overview}.

\begin{figure*}
  \centering
  \begin{minipage}[t]{0.48\textwidth}
    \includegraphics[width=\linewidth]{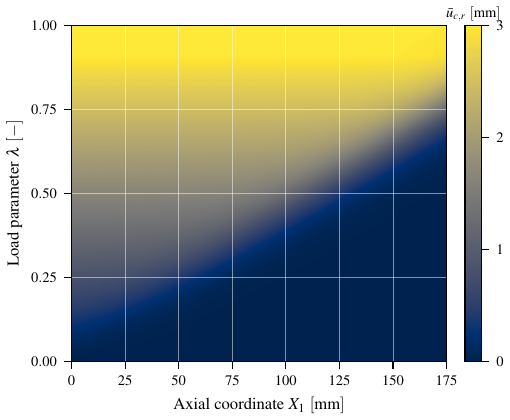}
    \caption{Inward radial displacement~$\radialdisplacementcompressiontest$ of the inner tube surface from the displacement
      function~$\displacementcompressiontest \klr{\xref, \loadstepparameter}$
      of~\cref{eq:crimpingdisplacementfunction} over the axial coordinate~$\xrefX$ and the load
      parameter~$\loadstepparameter$ for the parameters of the presented example.}
    \label{fig:crimpingdisplacementfunction}
  \end{minipage}
  \hfill
  \begin{minipage}[t]{0.48\textwidth}
    \includegraphics[width=\linewidth]{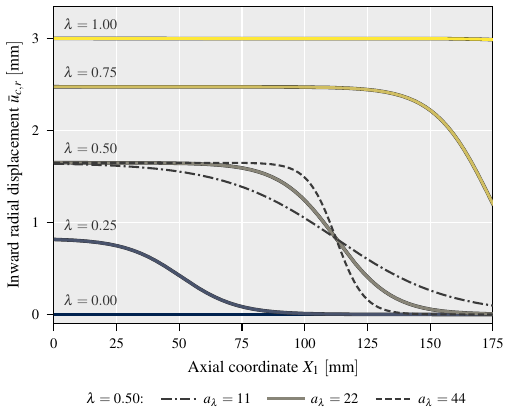}
    \caption{Inward radial displacement~$\radialdisplacementcompressiontest$ over~$\xrefX$
      for different values of~$\loadstepparameter$. With increasing~$\loadstepparameter$, the
      compression travels along the tube axis, and at~$\loadstepparameter=1$ the full
      displacement is reached along the entire tube.
      For~$\loadstepparameter=0.5$, two additional curves show the influence
      of~$\logisticsteepness$, each with~$\logisticaxialrate$ scaled by the same factor.}
    \label{fig:crimpingdisplacementcurves}
  \end{minipage}
\end{figure*}

A surrounding tubular structure with initial inner diameter of~$\dmcatinner =
\qty{18}{\milli\meter}$, a constant wall thickness of~$\mcatthickness =
\qty{0.1}{\milli\meter}$ and a length of~$\lmcat=\qty{175}{\milli\meter}$ is subjected
to a radial compression. As for the \gls{ac:FD} in~\cref{eq:fd_centerline}, the tube
axis is aligned with~$\unitvecx$, \ie~$\xrefX$ denotes the axial coordinate, measured
from the tube end at which the compression starts. The compression is formulated for
the inner tube surface, which is the contact surface of the \gls{ac:MC}. The
displacement field reads
\begin{equation}
  \begin{split}
    \displacementcompressiontest \klr{\xref, \loadstepparameter}
    ={}                                   &
    \frac{1}{
      1 + e^{\!\left(
          -\logisticsteepness \klr{\loadstepparameter-\loadstepshift}
          + \logisticaxialrate \xrefX
          \right)}
    }
    \\
                                          & \cdot
    \left(
    \frac{\targetradius(\loadstepparameter)}
    {\sqrt{\xrefY^2+\xrefZ^2}} - 1
    \right)
    \begin{bmatrix}
      0      \\
      \xrefY \\
      \xrefZ
    \end{bmatrix}
    \\
    \text{with~}
    \targetradius(\loadstepparameter) ={} & \tfrac{1}{2}\klr{\dmcatinner + \min\klr{\loadstepparameter, 1}\klr{\dmcatinnercompressed - \dmcatinner}}
    \text{.}
  \end{split}
  \label{eq:crimpingdisplacementfunction}
\end{equation}
The displacement field~\(\displacementcompressiontest\klr{\xref,\loadstepparameter}\) moves each point of
the inner surface towards the target radius~$\targetradius(\loadstepparameter)$, which
decreases linearly with~$\loadstepparameter$ until the compressed inner
diameter~$\dmcatinnercompressed=\qty{12}{\milli\meter}$ is reached at~$\loadstepparameter=1$, and
is held constant afterwards.
Since~$\displacementcompressiontest$ has no axial component and points towards the tube
axis, its magnitude~$\radialdisplacementcompressiontest=\norm{\displacementcompressiontest}$
is the inward radial displacement.
The applied displacement function ensures
a smooth and gradual compression along the axis of the \gls{ac:FD}.
The first factor of~\cref{eq:crimpingdisplacementfunction} is a logistic function
with respect to~$\xrefX$, whose transition between $1$ and $0$ is centered
at~$\xrefX=\logisticsteepness\klr{\loadstepparameter-\loadstepshift}/\logisticaxialrate$ and
travels along the tube axis with increasing~$\loadstepparameter$, starting at~$\xrefX=0$,
see~\cref{fig:crimpingdisplacementfunction,fig:crimpingdisplacementcurves}.
For the presented example, the characteristic parameters are selected
as~$\logisticsteepness=22$, $\loadstepshift=1/22$
and~$\logisticaxialrate=\tfrac{20}{225}\,\unit{\per\milli\meter}$, such that contact
between the \gls{ac:FD} and the \gls{ac:MC} develops gradually. In contrast to a purely
uniform radial compression applied per load step, this approach mitigates numerical
instabilities caused by the abrupt nonlinear effect of contact and thus significantly
improves the robustness of the example. Since the transition is centered
at~$\xrefX=\logisticsteepness\klr{\loadstepparameter-\loadstepshift}/\logisticaxialrate$,
scaling~$\logisticsteepness$ and~$\logisticaxialrate$ by the same factor keeps its
position and only changes its width. A larger~$\logisticsteepness$ yields a sharper
transition, see~\cref{fig:crimpingdisplacementcurves}.

On the inner surface of the tube, the beam-to-surface interaction
boundary~$\btscouplingcondition$ is considered, allowing the device to deform and
compress as a consequence. Each wire is subject to both beam-to-beam contact and a
beam-to-surface interaction condition, thereby capturing the interactions among the
individual wires as well as those with the inner surface of the cylinder. The boundary
condition indicated by the blue circular marker
in~\cref{fig:compressesion-test-overview} at the left end of the \gls{ac:FD} denotes
the introduction of local coordinate systems. The displacement components in the local
direction~$\rotunitlocx$ and~$\rotunitlocy$ are constrained, such that the wire end
point can only displace in radial direction.
In addition to the local coordinate system, the coincident end-node pairs marked in
blue are subjected to positional coupling. At the opposite end of the \gls{ac:FD}, the
red box-shaped marker denotes coincident end-node pairs that are subject to positional
coupling as well. All associated pairs are collected in the coupling
set~$\couplingset$. While~\cref{eq:crimpingdisplacementfunction} prescribes all three
displacement components of the tube, with a vanishing axial component, the axial
displacement of the \gls{ac:FD} is constrained only at its front face.

The geometric parameters of the~\gls{ac:FD} are identical to~\cref{sec:tensile-example}
and are summarized in~\cref{tab:example-01-parameters}. Each wire centerline of the
\gls{ac:FD} is discretized using~$101$ beam elements, with the Young’s
modulus~$\beamyoungsmod = \qty{206}{\giga\pascal}$ and shear
modulus~$\beamshearsmod=\qty{81.5}{\giga\pascal}$.

\begin{figure*}
  \centering
  \includegraphics[width=\linewidth]{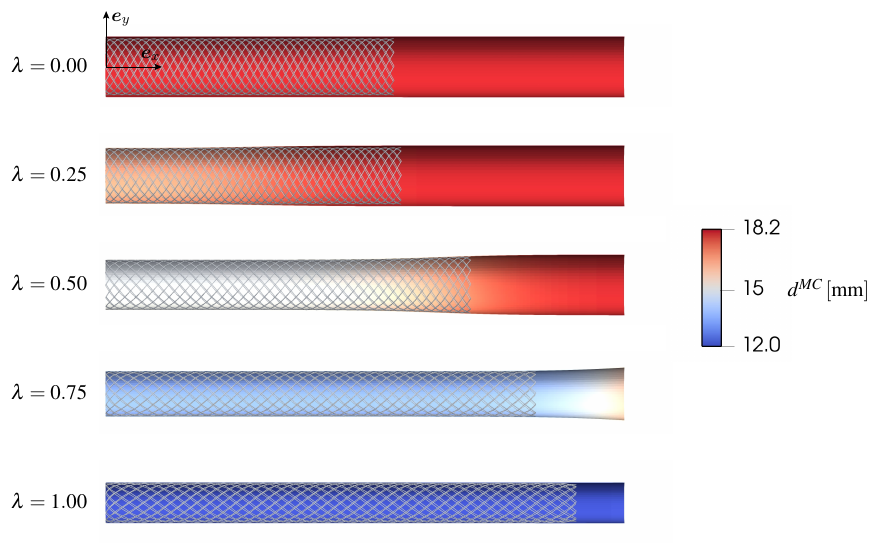}

  \caption{Visualization of the deformed configuration of a \gls{ac:FD} during the compression test.
    The figures present the solution at five values of the load parameter, from the initial state
    at~$\loadstepparameter=0$ to the final state at~$\loadstepparameter=1$.
    The compression of the solid tube extends continuously along the axis,
    and slowly establishes contact to the individual wires.
    The color bar indicates the progress of the compressed state with the current diameter~$\dmcat$
    of the tube. }
  \label{fig:radial-results}
\end{figure*}

\begin{figure}
  \begin{minipage}[b]{\linewidth}
    \centering
    \includegraphics[width=\linewidth]{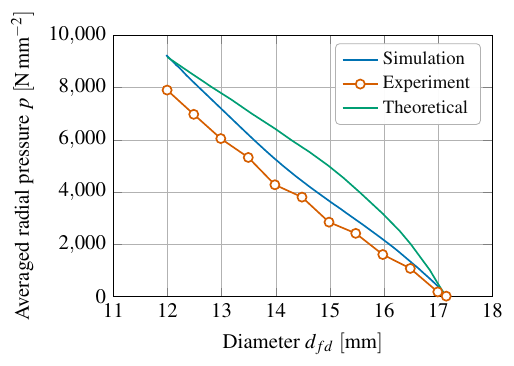}
  \end{minipage}
  \caption{Comparison of the evolving pressure during the crimping test case with changing diameter~$\dfd$.
    Experimental and theoretical reference data are taken from~\cite{jedwabStudyGeometricalMechanical1993}.}
  \label{fig:radial-test-results}
\end{figure}

Although the beam-to-surface contact interaction is evaluated at projected contact
points during the numerical solution procedure, the resulting contact forces are
consistently assembled into the global residual vector and are therefore represented at
the nodes. Consequently, the nodal force quantities obtained from the simulation
constitute the discrete representation of the total contact interaction within the
finite element formulation. In a post-processing step, the in-plane radial force
magnitude is evaluated at each node and summed up, resulting in a scalar measure for
the total radial contact force. This procedure ensures that all contact contributions
are accounted for, independent of the specific locations of the underlying contact
points. The resulting force is divided by the surface of the \gls{ac:MC} to compute the
presented average radial pressure $\pressure$, yielding a measure, which can be
directly compared to experimentally and theoretically observations
from~\cite{jedwabStudyGeometricalMechanical1993}.

The numerical results capture the characteristic mechanical response of a braided
\gls{ac:FD} under radial compression. As shown in~\cref{fig:radial-test-results}, the
averaged radial pressure increases monotonically with decreasing device diameter. The
numerical simulation results show in general good agreement with the selected reference
data and are bound between the theoretical predictions and the experimental
measurements.
For the sake of completeness, it is noted that a similar strain relation as within the
previous example is considered and calculated according
to~\cref{eq:strain-comparison-measure}. The obtained strain can be stated for this
example as~$\strainmax = \num{1.0533e-4}$. Therefore, the strain remains as well within
the small-strain regime, indicating that nonlinear material effects associated with
superelastic Nitinol do not have to be accounted for.

Overall, these results demonstrate that the proposed modeling approach reliably
reproduces the qualitative pressure--diameter relationship relevant for deployment and
placement simulations, while quantitative discrepancies arise from differences in
geometry and modeling assumptions.


\subsection{Geometrical test case}\label{sec:shapiro-test}
Apart from the overall structural response and the force response, the precise
geometric configuration of the individual wires is of particular interest. Especially
in the context of virtual fast stenting approaches and subsequent \gls{ac:CFD}
analyses, the final spatial configuration of the individual wires can be directly
incorporated into the fluid domain
geometry~\cite{STAHL2023106720,jeken-ricoVirtualFlowDiverter2024,Hagmeyer2022a,NoraHagmeyer}.
Alternatively, geometric parameters such as the resulting braiding angle and inter-wire
distance can be further processed to derive application-relevant quantities, including
the device porosity and the \gls{ac:MCR}. These metrics serve as important reference
measures in comparative studies~\cite{beppuComputationalFluidDynamics2020,
dholakiaHemodynamicsFlowDiverters2017} and provide a basis for further investigations
into device design and long-term prediction.

To provide further insight into these metrics and relate them to the proposed modeling
approach, the following experiment is considered.
Within~\cite{shapiroVariablePorosityPipeline2014}, an experimental study was conducted
in which a~\gls{ac:PED} was subjected to stepwise compression to several prescribed
constant diameters. The resulting axial distribution of the~\gls{ac:MCR} was
reconstructed using multiple imaging techniques. Due to the systematic geometric
changes introduced by the stepwise compression, this experiment constitutes a suitable
benchmark for assessing the geometric response predicted by the proposed framework.
Hence, within the scope of this example, it will be outlined how such a stepwise
compression of a \gls{ac:FD} can be performed. In the following, the changes in
braiding angle and distance between wires along the device are analyzed and linked to
the resulting \gls{ac:MCR}. Since the exact values of the braiding
angle~$\fdbraidingangle$ were not reported, additional variations of~$\pm
\qty{4}{\degree}$ with respect to the known reference values of a~\gls{ac:PED} are
considered for the analysis of the \gls{ac:MCR}.

To this end, the example introduced in~\cref{sec:crimping-test} is extended by
subjecting the device to a prescribed sequence of radial compression steps. The basic
setup of the example with respect to boundary conditions is identical to the example
displayed in~\cref{fig:compressesion-test-overview}. However, the geometric quantities
of the \gls{ac:FD} and \gls{ac:MC} differ. The considered~\gls{ac:FD} is a~\gls{ac:PED}
with a diameter~$\dfd=\qty{4.25}{\milli\meter}$, a length
of~$\lfd=\qty{20}{\milli\meter}$ and a helix angle
of~$\fdhelixangle=\qty{68.8}{\degree}$. The respective wire radius of a single wire
is~$\rfdwire=\qty{15}{\micro\meter}$ and the distance between two wires is given
as~$\fdw=\qty{0.2959}{\milli\meter}$~\cite{bouillotGeometricalDeploymentBraided2016}.
The \gls{ac:FD} is created from the parametric equations presented
in~\cref{sec:fdparametric} with an interwoven amplitude equal to the wire radius and a
single crossing braiding pattern~$\fdinterwoovenfamiliy=1$. Each wire is discretized
with~$299$ elements, ensuring that most of the initial contact points between the beams
are located within the center of a beam element. The beam Young’s
modulus~$\beamyoungsmod$ is assumed to be~$\qty{83}{\giga\pascal}$ and shear
modulus~$\beamshearsmod=\qty{41.5}{\giga\pascal}$. The point contact penalty
parameter~$\btbpointpenalty$ is increased linearly with respect to the load
parameter~$\loadstepparameter$ starting from~$1$ to~$\num{8500}$.
The tube has a length of~$\lmcat=\qty{34}{\milli\meter}$ and an initial inner diameter
of~$\dmcatinner=\qty{4.5}{\milli\meter}$, corresponding to the initial inner
radius~$\rmcatinner=\qty{2.25}{\milli\meter}$ and a thickness
of~$\mcatthickness=\qty{0.1}{\milli\meter}$.

\begin{figure*}
  \centering
  \begin{minipage}[b]{0.45\textwidth}
    \includegraphics[width=\linewidth]{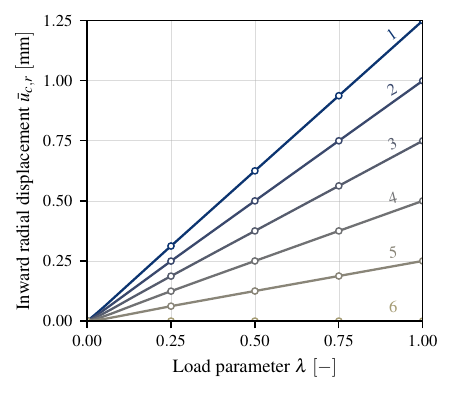}
    \caption{Visualization of the inward radial displacement~$\radialdisplacementcompressiontest$ for each center of the segments over the load parameter. }\label{fig:segmentwise-real-displacement-over-lambda}

  \end{minipage}\hspace{2em}
  \begin{minipage}[b]{0.45\textwidth}
    \includegraphics[width=\linewidth]{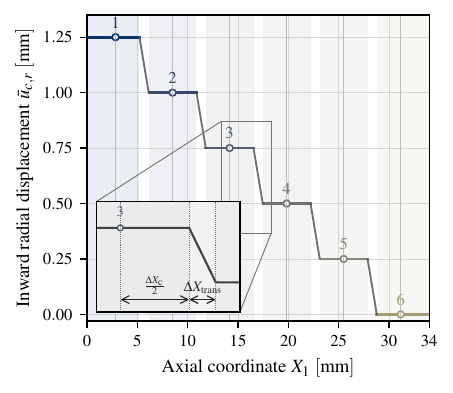}
    \caption{Visualization of the inward radial displacement~$\radialdisplacementcompressiontest$ at $\loadstepparameter=1$ over the whole axial length. \\}\label{fig:dislacment-over-axial-lenght}
  \end{minipage}
\end{figure*}

The final compressed configuration is defined by six equally sized axial segments. For
each segment~$\segmentindex$, the final target inner radius of the \gls{ac:MC}, denoted
by~$\segmenttargetradius$, is prescribed according to the stepwise radius variation
from the experiment. The corresponding values are listed
in~\cref{tab:stepwise-target-radii}. Within each axial segment, the target radius is
kept constant over an interval of width~\(\segmentwidth=\qty{4.78125}{\milli\meter}\)
centered at the respective segment center. Between neighboring segments, a transition
region of width~\(\segmenttransitionwidth=\qty{1.0625}{\milli\meter}\) is introduced,
in which the target radii of the two adjacent segments are linearly interpolated. This
interpolation defines a continuous target-radius field
\(\targetradius(\xrefX,\loadstepparameter)\) with
\(\targetradius(\xrefX,1)=\segmenttargetradius\) within segment~$\segmentindex$,
avoiding discontinuities in the prescribed displacement field while retaining the
intended stepwise character of the compression, which can be seen
in~\cref{fig:dislacment-over-axial-lenght}.

\begin{figure*}
  \centering
  \begin{minipage}[b]{\textwidth}
    \centering
    \includegraphics[width=\linewidth]{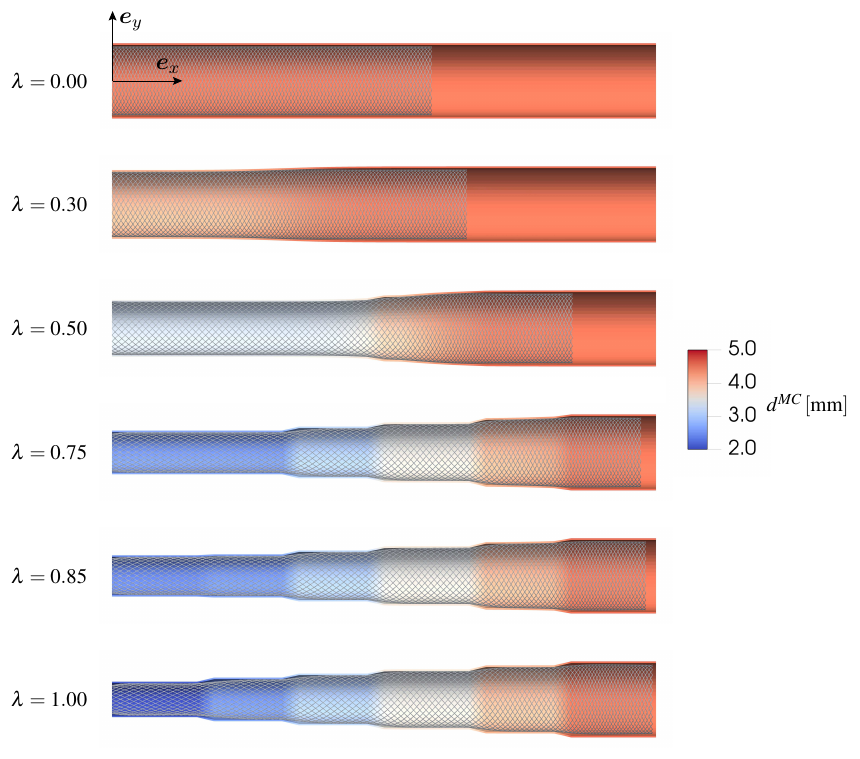}
  \end{minipage}

  \caption{Visualization of the stepwise deformation for the \gls{ac:FD} with a braiding angle of~$\fdbraidingangle=\qty{137.6}{\degree}$.
    The figures present the solution at six values of the load parameter, from the initial state
    at~$\loadstepparameter=0$ to the final state at~$\loadstepparameter=1$ with respect to the changing diameter of the tube~$\diameter^{\mcat}$.}
  \label{fig:step-results}
\end{figure*}

\begin{table}[h]
  \centering
  \begin{tabular}{c c c}
    \hline
    Segment                & Center position
                           & Target radius                                          \\
    Number~$\segmentindex$ & ~$\segmentcenter$~$[\unit{\milli\meter}]$
                           & ~$\segmenttargetradius$~$[\unit{\milli\meter}]$
    \\
    \hline
    1                      & 2.83                                            & 1.00 \\
    2                      & 8.50                                            & 1.25 \\
    3                      & 14.17                                           & 1.50 \\
    4                      & 19.83                                           & 1.75 \\
    5                      & 25.50                                           & 2.00 \\
    6                      & 31.17                                           & 2.25 \\
    \hline
  \end{tabular}
  \caption{Segment numbers with center positions and prescribed inner target radii used for the
    stepwise compression.}
  \label{tab:stepwise-target-radii}
\end{table}
The following displacement prescription is formulated for the inner tube
surface. The outer surface and the remaining tube volume can be treated
analogously by defining the corresponding target-radius field. The displacement
field is prescribed such that each point is moved from its initial radial
position towards the interpolated target radius and is given by
\begin{equation}
  \begin{split}
    \displacementcompressiontest(\xref,\loadstepparameter)
    ={} &
    \frac{1}{
      1 + e^{\!\left(
          -\logisticsteepness \klr{\loadstepparameter-\loadstepshift}
          + \logisticaxialrate \xrefX
          \right)}
    }
    \\
        & \cdot
    \left(
    \frac{\targetradius(\xrefX,\loadstepparameter)}
    {\sqrt{\xrefY^2+\xrefZ^2}} - 1
    \right)
    \begin{bmatrix}
      0      \\
      \xrefY \\
      \xrefZ
    \end{bmatrix}.
  \end{split}
  \label{eq:stepwisecompressionfct}
\end{equation}
The first factor in~\cref{eq:stepwisecompressionfct} defines a smooth activation of the
compression along the tube axis, with~\(\logisticsteepness=30\),
\(\loadstepshift=1/150\), and~\(\logisticaxialrate=20/\lmcat\). The second factor
scales the radial position of each constrained point to the prescribed target radius.
The resulting displacement profile is illustrated
in~\cref{fig:segmentwise-real-displacement-over-lambda,fig:dislacment-over-axial-lenght}.
In~\cref{fig:step-results} the simulation results are shown.

\begin{figure}
  \centering
  \begin{minipage}[b]{\linewidth}
    \includegraphics[width=\linewidth]{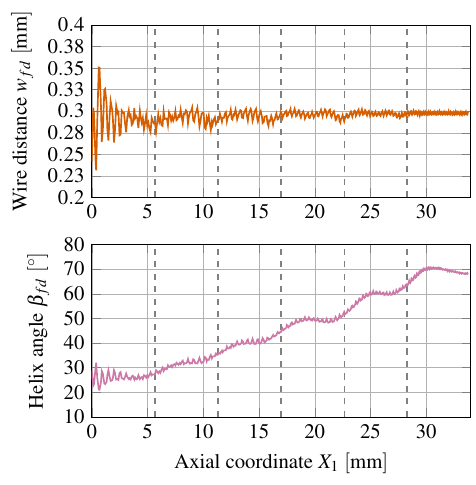}
    \caption{Visualization of the evolving
      distance between two selected wires and of the helix angle over the axial length
      for the \gls{ac:FD} with a helix angle of~$\fdhelixangle=\qty{68.8}{\degree}$.
      The dashed lines indicate the six axial segments resulting from the stepwise compression.}
    \label{fig:step-compression-width-angle}
  \end{minipage}
\end{figure}

Within a post-processing step the distance~$\fdw$ from~\cref{fig:realfd} is obtained
from two selected wires with the same clock-wise orientation. The helix
angle~$\fdhelixangle$ is recalculated from the beam tangent vector~$\beamtangentvector$
relative to the unit vector~$\unitvecx$, which is aligned with the cylindrical axis.
The resulting post-processed quantities are shown
in~\cref{fig:step-compression-width-angle}. Within each compressed segment, the helix
angle approaches an approximately constant level, with small local deviations caused by
the explicitly resolved interwoven wire geometry. The inter-wire distance~$\fdw$ shows
a less pronounced plateau structure, but its deviation from the initial value increases
with stronger radial compression. At the interfaces between adjacent axial segments,
small but visible variations in~$\fdw$ occur. The local width variations are relevant
when comparing against simplified geometric deployment approaches, where similar
effects may depend on the underlying geometric assumptions. Due to the symmetry of the
example, the helix angle displayed in~\cref{fig:step-compression-width-angle} can be
related to the contact angle. The comparatively large contact angles observed
throughout the simulation support the use of the point-contact contribution for the
present example.

\begin{figure}
  \centering
  \begin{minipage}[b]{\linewidth}
    \includegraphics[page=1,width=\linewidth]{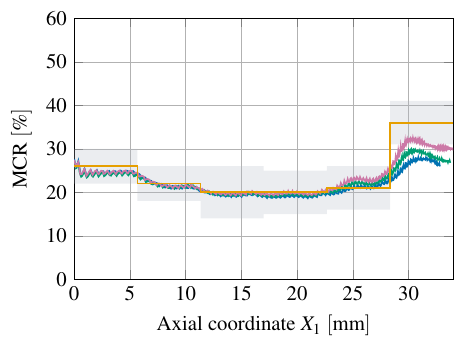}\\[1ex]
    \centering
    \includegraphics[page=2,scale=1.08]{results-step-compression/mcr_diff_angles.pdf}\vspace{1em}
    \caption{Resulting values for the \gls{ac:MCR} for different braiding angles~$\fdbraidingangle$ displayed over the device length with varying diameters.
      Reference values from~\cite{shapiroVariablePorosityPipeline2014} are included with the reference range.}
    \label{fig:step-compression-mcr}
  \end{minipage}
\end{figure}
Based on the helix angle~$\fdhelixangle$ and the inter-wire distance~$\fdw$, the \gls{ac:MCR}, as defined
in~\cite{shapiroVariablePorosityPipeline2014,jeken-ricoVirtualFlowDiverter2024,bouillotGeometricalDeploymentBraided2016},
can be recalculated starting from the local porosity~$\mcrporositiy$ according to
\begin{equation}
  \begin{split}
    \mcrporositiy\!\left(\fdhelixangle\right)
         & =
    \klr{1 -
      \frac{2\rfdwire}{\fdw \sin\!\left(2\fdhelixangle\right)}}^2\quad\text{~and~}
    \\
    \mcr & = 100\klr{1-\mcrporositiy}\text{.}
  \end{split}
  \label{eq:mcr}
\end{equation}
The local porosity~$\mcrporositiy\!\left(\fdhelixangle\right)$ is derived by considering a rhombic unit cell formed
by four intersecting wires, with the braiding angle~$\fdbraidingangle$ at the crossing points of the wires.
In~\cref{fig:step-compression-mcr}, the \gls{ac:MCR} is presented for
three devices with different braiding angles.
According to the analysis by~\cite{bouillotGeometricalDeploymentBraided2016}, the \gls{ac:MCR} reaches a local
minimum around~$\qty{45}{\degree}$, which for the presented case is within the third and
fourth segment. For each device visualized in~\cref{fig:step-compression-width-angle},
the \gls{ac:MCR} remains nearly constant.
Overall the devices match the given target range from the reference~\cite{shapiroVariablePorosityPipeline2014} very well.
The largest difference can be observed, within the last segment, where the initial~\gls{ac:MCR}
is recovered.
With the proposed braiding angle of~$\fdbraidingangle=\qty{137.6}{\degree}$,
the initial \gls{ac:MCR} is approximately~\qty{27}{\percent}, which is below the reference range of~\qtyrange{31}{41}{\percent}.
Therefore, two small variations of the braiding angle are examined.
Decreasing the braiding angle to~$\fdbraidingangle=\qty{133.6}{\degree}$ results
in a lower \gls{ac:MCR} and increasing the braiding angle to a value
of~$\fdbraidingangle=\qty{141.6}{\degree}$, results in a higher \gls{ac:MCR}.
For the increased braiding angle~$\fdbraidingangle=\qty{141.6}{\degree}$
the \gls{ac:MCR} tends now towards the lower boundary of
the reference range.
Due to the change within the braiding angle, the respective length of the
\glspl{ac:FD} change, which explains the missing \gls{ac:MCR} towards the axial end position.
Since the braiding angle is considered to be a key design variable, and given the time elapsed between the reference data reported~\cite{bouillotGeometricalDeploymentBraided2016}
and those reported in~\cite{shapiroVariablePorosityPipeline2014}, a plausible reason can be
that the design of the \gls{ac:FD} was slightly adapted and therefore, this discrepancy within
\gls{ac:MCR} can be observed. Another aspect could be the missing
friction between the \gls{ac:FD} and \gls{ac:MC}, since an increasing value of
$\fdbraidingangle$ would result, according to the definition from~\cref{eq:mcr}, into a higher \gls{ac:MCR}.

\begin{figure*}
  \centering

  \begin{minipage}{0.49\linewidth}
    \centering
    \includegraphics[width=\linewidth]
    {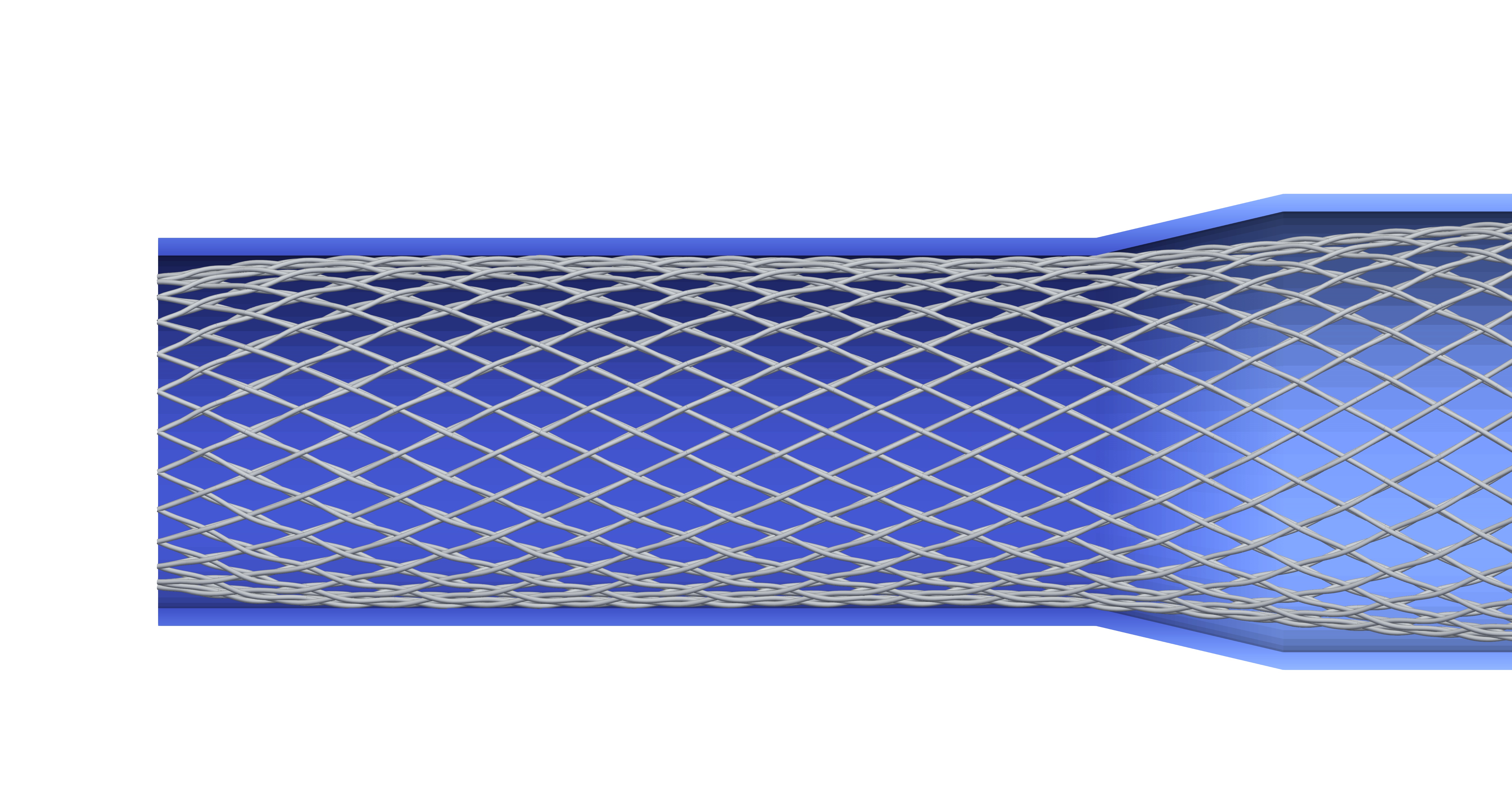}
  \end{minipage}
  \hfill
  \begin{minipage}{0.49\linewidth}
    \centering
    \includegraphics[width=\linewidth]
    {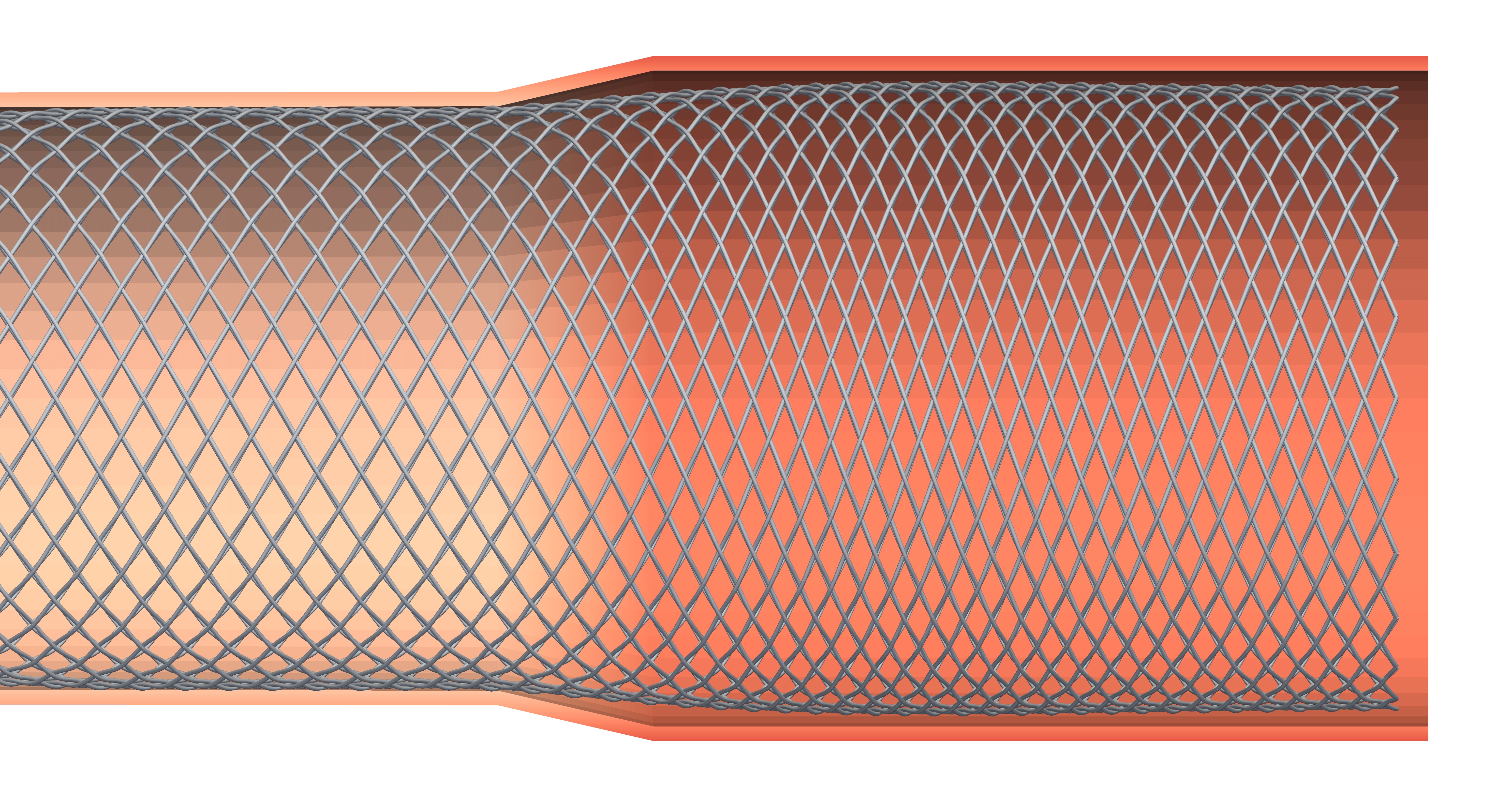}
  \end{minipage}

  \caption{Visualization of the ends of the \gls{ac:FD} in the first and last segment
    for a braiding angle of~$\fdbraidingangle=\qty{137.6}{\degree}$.}
  \label{fig:step-compression-both-ends}
\end{figure*}

\begin{figure}
  \centering

  \includegraphics[width=\linewidth,
    trim=0 0 0 450, clip]
  {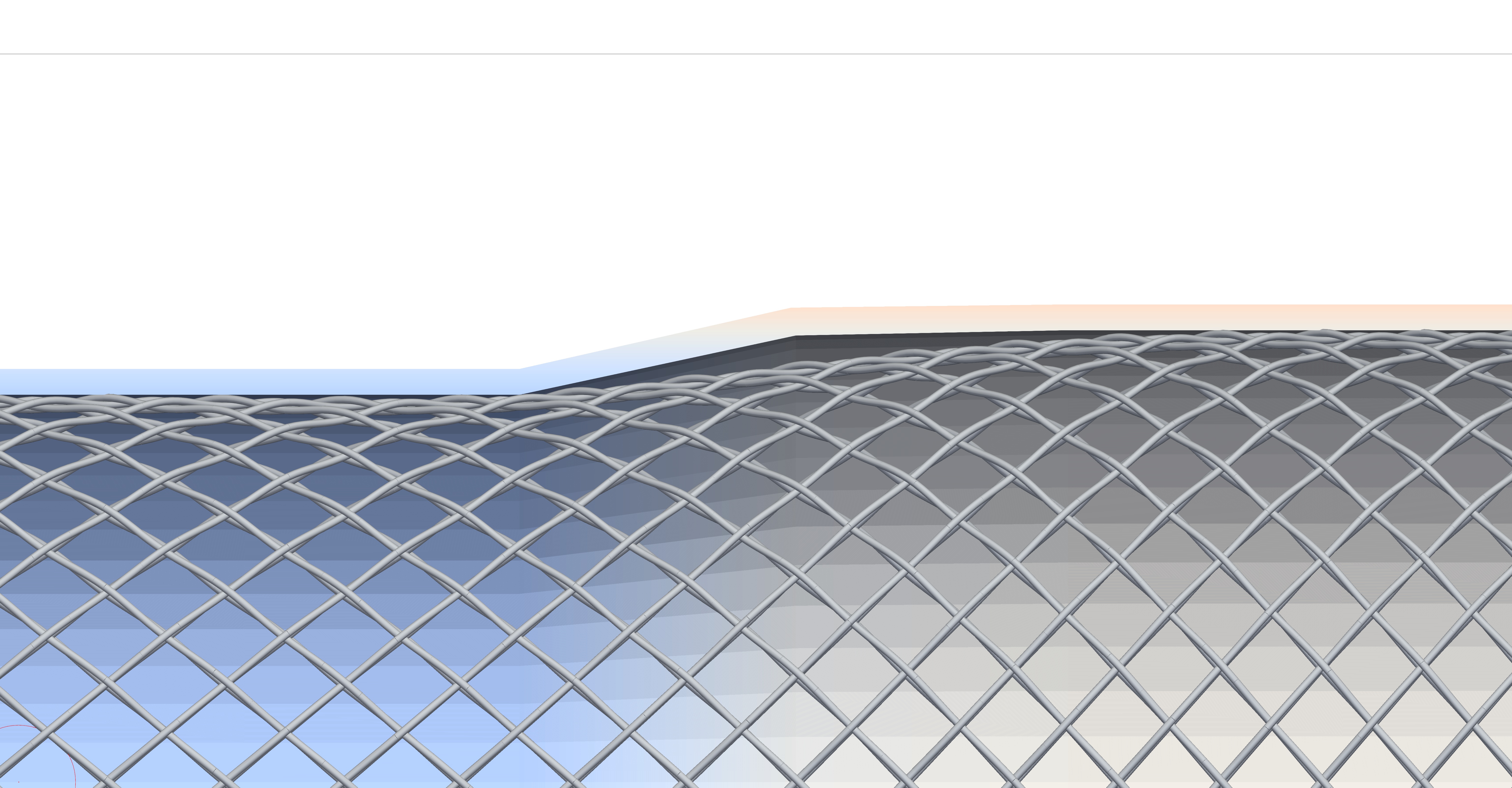}
  \caption{Detailed view of the braiding pattern of a \gls{ac:FD} with a braiding angle of
    $\fdbraidingangle=\qty{137.6}{\degree}$, showing the smooth transition between the third
    and fourth segment of the \gls{ac:MC}.}
  \label{fig:step-compression-detail-step}

\end{figure}

In~\cref{fig:step-compression-both-ends}, both end regions of the device for the
presented modeling approach are depicted. At the smaller-diameter side, the influence
of the positional coupling can be clearly observed, as the local diameter of the
\gls{ac:FD} reduces. At the larger-diameter side, a change in the helix angle is
visible, accompanied by a slight reduction in device diameter toward the end region as
a consequence of the applied coupling conditions. As shown
in~\cref{fig:step-compression-detail-step}, the beams align smoothly with the inner
wall of the tube for diameters of~$\qty{3.0}{\milli\meter}$
and~$\qty{3.5}{\milli\meter}$. The braiding pattern is fully preserved, and no
geometric intersections between wires are observed, indicating a physically consistent
and realistic deformation behavior of the \gls{ac:FD} within the case of a stepwise
radial compression.

The presented example further demonstrates a key advantage of the proposed approach.
Since each wire and its contact interactions are resolved, geometric effects induced by
the stepwise variation of the diameter are naturally captured, in contrast to
simplified virtual stenting techniques~\cite{jeken-ricoVirtualFlowDiverter2024}. In
particular, as shown in~\cref{fig:step-compression-detail-step}, the transition between
regions of smaller and larger diameter is resolved smoothly by the individual wires,
reflecting the underlying braid mechanics.

\section{Conclusion}\label{sec:conclusion}

In this work, a finite element--based modeling approach for the mechanical simulation
of braided neurovascular \glspl{ac:FD} has been developed. The model integrates a
geometrically exact Simo--Reissner beam formulation with beam-to-beam and frictionless
beam-to-surface contact to capture the interactions between the wires of a \gls{ac:FD}
and its contact interaction with a \gls{ac:MC}. A parametric description of interwoven
braided \glspl{ac:FD} was introduced, enabling the generation of different braiding
patterns in a geometrically consistent manner. Beam-to-beam contact ensures the
kinematic consistency of the braided structure, while beam-to-surface contact captures
the interaction with surrounding tubular structures such as \gls{ac:MC}s or crimping
devices.

The framework was compared against three representative test scenarios from the
literature. In a uniaxial tensile test, the model reproduced the characteristic
length--diameter relationship and axial force response with good agreement. In the
crimping scenario, the predicted radial pressure evolution was found to lie between
theoretical and experimental reference values, demonstrating physically meaningful
contact behavior. Finally, the stepwise compression example confirmed the capability of
the framework to capture geometric measures such as wire distance, helix angle
variation, and the \gls{ac:MCR}. The computed \gls{ac:MCR} values showed good agreement
with the reported reference ranges, highlighting the suitability of the approach for
geometry-sensitive deployment analyses.

Although a linear elastic constitutive model was used for the individual wires, the
observed strain levels in all investigated examples remained within the small-strain
regime, which justifies this assumption for the present validation cases. Future work
could therefore focus on incorporating nonlinear, phase-transforming superelastic
material models to capture cyclic loading effects and thermal dependencies more
realistically. In addition, the current validation cases are restricted to straight
configurations, whereas clinical deployment often involves curved and tortuous
anatomies. Therefore, the behavior of the proposed model should also be assessed in
bending-dominated and patient-specific geometries. The present results nonetheless
provide a solid mechanical basis for such extensions and demonstrate the robustness of
the proposed formulation. Further developments may also include frictional contact or
global sensitivity analyses to assess the influence of key design parameters.

Overall, the proposed modeling framework provides a mechanically consistent and
geometrically detailed basis for structural and geometric analyses of braided
\glspl{ac:FD} and establishes a foundation for future patient-specific deployment
simulations and hemodynamic investigations.

\backmatter

\bmhead{Acknowledgements}
The figures presented in this work have been created using the Adobe Illustrator
plug-in LaTeX2AI \url{https://github.com/latex2ai/LaTeX2AI}.

\section*{Statements and Declarations}

\bmhead{Funding}
This work was funded by the Deutsche Forschungsgemeinschaft (DFG, German Research
Foundation) within the project ``Multi-scale algorithms and simulation methodologies for
the long-term prognosis of endovascular interventions in cerebral aneurysms'' (project
number 465242983) within the DFG priority programme ``SPP 2311: Robust coupling of
continuum-biomechanical in silico models to establish active biological system models
for later use in clinical applications -- Co-design of modeling, numerics and
usability''.

\bmhead{Competing interests}
The authors have no competing interests to declare that are relevant to the content of
this article.

\bmhead{Data availability}
All simulations were conducted using the open-source multiphysics solver \fourc. The
remaining data that support the findings of this study are available from the
corresponding author upon reasonable request.

\bmhead{Author contributions}
\textbf{Martin Frank:} Writing -- Original Draft Preparation, Visualization, Methodology, Software.
\textbf{Ivo Steinbrecher:} Writing -- Review \& Editing, Methodology, Software, Supervision.
\textbf{Matthias Mayr:} Writing -- Review \& Editing, Methodology, Project Administration, Supervision, Resources, Funding Acquisition.
\textbf{Alexander Popp:} Writing -- Review \& Editing, Resources, Funding Acquisition.
All authors read and approved the final manuscript.

\bmhead{Use of generative AI}
During the preparation of this work, the authors used \textit{OpenAI GPT-5.5} to improve
the readability and language of the manuscript. After using this tool, the authors
reviewed and edited the content as needed and take full responsibility for the content
of the published article.

\bibliography{bibliography}

\end{document}